\documentclass[amsmath,amssymb,twocolumn,amsfonts]{revtex4-2}

\def \beq {\begin{equation}}
\def \eeq {\end{equation}}
\def \ba {\begin{align}}
\def \ea {\end{align}}
\def \tr {\rm Tr}

\usepackage{epsfig}
\usepackage{amssymb}
\usepackage{mathrsfs}
\usepackage{bm}
\usepackage{bbm,mathbbol}
\usepackage{braket,bigints}
\usepackage{stmaryrd,mathtools}
\usepackage{graphicx}
\usepackage{dcolumn}
\usepackage{xcolor}
\usepackage{bbold}
\usepackage{xparse}
\usepackage{hyperref}
\usepackage{siunitx}
\usepackage{cleveref}
\usepackage{braket,bigints}
\usepackage[T1]{fontenc}

\newcommand{\Rb}{\text{Rb}}
\newcommand{\Cs}{\text{Cs}}
\renewcommand\bra[1]{{\langle{#1}|}}
\makeatletter
\renewcommand\ket[1]{%
\@ifnextchar\bra{\k@t{#1}\!}{\k@t{#1}}%
}
\newcommand\k@t[1]{{|{#1}\rangle}}
\makeatother

\usepackage{color}
\definecolor{mygreen}{rgb}{0,0.5,0}
\definecolor{mygrey}{rgb}{0.5,0.5,0.5}
\definecolor{myred}{rgb}{0.75,0,0}
\definecolor{myblue}{rgb}{0,0,0.75}
\definecolor{mymagenta}{cmyk}{0,1,0,0.12}
\definecolor{mycyan}{cmyk}{1,0,0,0.12}
\definecolor{myorange}{rgb}{1.,0.5,0}
\definecolor{myviolet}{rgb}{0.6,0.15,0.6}
\definecolor{mybrown}{cmyk}{0,0.50,1,0.41}

\usepackage{ulem}

\newcommand{\NAtoms}{N_\mathrm{at}}

\begin{document}
\title{Inter-species spin-noise correlations in hot atomic vapors}
\author{K. Mouloudakis\textsuperscript{1,2}}
\author{F. Vouzinas\textsuperscript{1}}
\author{A. Margaritakis\textsuperscript{1}}
\author{A. Koutsimpela\textsuperscript{3}}
\author{G. Mouloudakis\textsuperscript{1,4}}
\author{V. Koutrouli\textsuperscript{1,4}}
\author{M. Skotiniotis\textsuperscript{5,6}}
\author{G. P. Tsironis\textsuperscript{1,7}}
\author{M. Loulakis\textsuperscript{3,8}}
\author{M. W.  Mitchell\textsuperscript{2,9}}
\author{G. Vasilakis\textsuperscript{4}}
\email{gvasilak@iesl.forth.gr}
\author{I. K. Kominis\textsuperscript{1,7,10}}
\email{ikominis@uoc.gr}
\affiliation{
$^1$Department of Physics, University of Crete, Heraklion 71003, Greece\\
$^2$ICFO - Institut de Ci\`encies Fot\`oniques, The Barcelona Institute of Science and Technology, 08860 Castelldefels (Barcelona), Spain\\
$^3$School of Applied Mathematical and Physical Sciences, National Technical University of Athens, 15780 Athens, Greece\\
$^4$Institute of Electronic Structure and Laser, Foundation for Research and Technology, 71110 Heraklion, Greece\\
$^5$F\'{\i}sica Te\`{o}rica: Informaci\'{o} i Fen\`{o}mens Qu\`{a}ntics, Departament de F\'{\i}sica, Universitat Aut\`{o}noma de Barcelona, 08193 Bellaterra, Spain\\
$^6$Departamento de Electromagnetismo y F\'{i}sica de la Materia, Universidad de Granada, 18010 Granada, Spain\\
$^7$Institute of Theoretical and Computational Physics, University of Crete, Heraklion 70013, Greece\\
$^8$Institute of Applied and Computational Mathematics, Foundation for Research and Technology, 70013 Heraklion, Greece\\
$^9$ICREA - Instituci\'{o} Catalana de Recerca i Estudis Avan{\c{c}}ats, 08010 Barcelona, Spain\\
$^{10}$Quantum Biometronics, Heraklion 71409, Greece
}
\begin{abstract}
We report an experimental and theoretical study of spin noise correlations in a \textsuperscript{87}Rb-\textsuperscript{133}Cs unpolarized alkali-metal vapor dominated by spin-exchange collisions. We observe strong unequal-time inter-species correlations and account for these with a first-principles theoretical model. Since the two atomic species have different spin precession frequencies, the dual-species vapor enables the use of an additional experimental handle, the applied magnetic field, for untangling various sub-types of spin correlations. In particular, the measured cross-correlation and auto-correlation spectra shed light on a number of  spin-dynamic effects involving intra-atom, inter-atom, intra-species and inter-species correlations. Cross-correlation coefficients exceeding $60\%$ have been observed at low-magnetic fields, where the two spin species couple strongly via spin-exchange collisions. The understanding of such spontaneously generated correlations can motivate the design of quantum-enhanced measurements with single or multi-species spin-polarized alkali-metal vapors used in quantum sensing applications.\end{abstract}

\maketitle

\section{Introduction}

Quantum measurements are at the basis of quantum technologies, from atomic magnetometers \cite{BudkerRomalis}, atomic clocks \cite{RevModPhys.87.637} and quantum optical measurements \cite{Bouwmeester1997}, to quantum computers \cite{Kielpinski2002}, quantum simulators \cite{PhysRevLett.124.110503,PRXQuantum.1.020308}, and even the detection of gravitational waves \cite{Aasi20131,PhysRevLett.124.171102}.  Quantum uncertainty and its dynamic manifestation imposes limitations to the precision of quantum measurements \cite{gardiner2004quantum}. Thus a major effort of modern quantum technology has been to engineer quantum noise, for example by generating squeezed states of light or atoms, in order to surpass what are broadly known as the standard quantum limits to measurement precision \cite{Braginsky1992}. These are limits usually applied to many-body systems, following under the working assumption of the particles being in separable quantum states, i.e. sharing no correlations. Yet, correlations are at the heart of the second quantum revolution \cite{PRXQuantum.1.020101}. Clearly, engineering quantum noise towards advancing the capabilities of quantum technology requires a  profound understanding of the physics of noise, taking correlations into account \cite{adesso2016measures}. 

Here we unravel multifaceted spin correlations spontaneously generated in a dual-species hot alkali-metal vapor. Hot atomic vapors are instrumental in quantum sensing of magnetic fields \cite{optical_mag}, as well as in atomic vapor clocks, while noble gas ensembles have been recently used for implementing quantum information protocols \cite{Ofer1,Ofer2,Shaham2022}, including quantum memories \cite{OferScience}. Moreover, numerous quantum systems can be mapped to spin, hence spin-noise studies in atomic vapors find direct analogies to other quantum technologies \cite{kopciuch2023optimized,Kitching_Chip_scale}. 

We present a precision spin-noise measurement in an unpolarized \textsuperscript{87}\rm{Rb}-\textsuperscript{133}\rm{Cs}  alkali-metal ensemble dominated by spin-exchange interactions. We demonstrate strong inter-species correlations at low magnetic fields, which fade away at increasing magnetic fields, as the two atomic species have different gyromagnetic ratios. The developed theoretical framework takes advantage of the long-standing physical description of spin-exchange collisions \cite{Happer_Book}, and leads to a formal description of spin-noise correlations, showing excellent agreement with the measurements. In particular, the magnetic-field dependence of the correlations in conjunction with the theoretical framework allows to resolve intra-atom from inter-atom correlations, and for each type discern intra-hyperfine from inter-hyperfine correlations.

This work has direct ramifications for understanding quantum limits to sensing technologies. This is because the observed correlations are created spontaneously by the ubiquitous spin-exchange collisions. Hence, what might have been understood as a vapor consisting of uncorrelated atoms {\it in the absence} of external perturbations, is actually an atomic vapor rich in correlations. These directly influence spin-noise benchmarks against which any noise engineering protocols which {\it do involve} external perturbations have to be compared. 

Moreover, while this measurement is performed with unpolarized atomic vapors, the developed theoretical framework allows us to extrapolate to polarized vapors pertinent to quantum sensing applications. In this regime we expect strong inter-species quantum correlations, which could further advance quantum metrology with multi-species hot atomic vapors.
 
The structure of the paper is the following. In the next section we provide a detailed perspective of this work in the context of previous work on quantum sensing with hot atomic vapors and spin-noise spectroscopy. In Sec. \ref{sec:exp} we describe the experimental setup and define the measured observables. In Sec. \ref{sec:theory} we present a comprehensive theoretical analysis of spin-noise correlations. We then analyze the data in Sec. \ref{sec:exp_analysis}, present the main results in Sec. \ref{sec:results}, and conclude with Sec. \ref{sec:conc}. Technical derivations are left for the appendices \ref{sec:AppA}-\ref{sec:app:BandwidthEffect}. 
 \section{\label{sec:sec2}This work in the context of previous work}
 Composite quantum systems, like a collection of interacting atoms or molecules and their interface with light, have been widely used as a realization of quantum sensing technologies \cite{PhysRevLett.111.120401,Degen,MacFarlane2003RSTA361,len2022quantum,amoros2021noisy}. In particular, hot alkali-metal vapors form the core quantum system in optical magnetometry \cite{PhysRevLett.109.253605, Koschorreck,mitchell2019quantum,PhysRevA.74.063420,Sheng_PRApplied_18_2022}, comagnetometry \cite{RomalisVas-comag,lee2022laboratory,wei2022ultrasensitive,wei2023dark}, magnetic field gradiometry \cite{PhysRevApplied.18.L021001,PhysRevApplied.15.014004}, frequency standards \cite{Almat:18} and quantum communications \cite{OferScience}. The long coherence times in combination with technical advantages, like optical accessibility with commercially available resonant laser light, and the high reliability of accommodating experimental setups, render hot alkali vapors  favorable in many quantum sensing applications, including magnetoencephalography \cite{PhysRevApplied.14.011002,Boto}, time keeping \cite{kang2015demonstration}, inertial sensing \cite{PhysRevA.94.032109,FangSensors2012} and imaging (THz imaging, biomagnetic imaging) \cite{PhysRevX.10.011027,Chen:22}.

Spin-exchange (SE) collisions, deriving from the Pauli exchange interaction during binary atomic encounters, are central in the physics of hot alkali vapors. The early understanding of SE collisions  \cite{Grossetete1,Grossetete2} accounted for several experimental observables \cite{PhysRevA.77.033408}. Comprehension of more subtle aspects of SE  \cite{PhysRevLett.31.273,PhysRevA.16.1877,HapperRev} led to the development of SERF (spin-exchange-relaxation-free) magnetometers \cite{Kominis2003,Allred,PhysRevA.80.013416}, SE optically-pumped hyperpolarized noble gases utilized in medical applications \cite{RevModPhys.69.629,walker2016spin,PhysRevA.42.1293}, and hydrogen magnetometry \cite{dikopoltsev2022magnetic,sofikitis2018ultrahigh,D1CP03171F}.
 
Rarely did the concept of correlations appear in these works, due to the intuitive expectation that, in hot vapors dominated by random binary spin-dependent collisions, correlations can be hardly sustained for meaningful time-scales \cite{AuzinshPRL}. However, several works have recently made such a counter-intuitive case on the non-trivial role of correlations, pointing to the possibility of performing quantum-enhanced measurements with hot vapors \cite{KuzmichSpinSqueezingQND,Julsgaard2001,Sherson2006,Kominis2008,VasilakisPolzik2015,MoulouPRR,MitchellNatureCom,PhysRevA.103.L010401,TrutleinHe3,Troullinou}. 
\subsection{Spin-noise spectroscopy}
In such studies, unpolarized alkali-metal vapors are rather useful, because unlike spin-polarized states, they are not sensitive to technical (e.g. magnetic) noise, hence intrinsic spin fluctuations can be readily measured. Since correlations and fluctuations are intimately related, as will be elaborated in detail in this work, unpolarized vapors are a natural testbed for studying atomic correlations. 

In more detail, unpolarized vapors provide easy access to the spontaneous fluctuations of atomic spin driven by atomic collisions. The field studying such fluctuations, spin-noise spectroscopy (SNS) \cite{Bloch,aleksandrovzapasskii,crooker2004,katsoprinakis,Mihailia,SinitsynSN,ZapasskiiSN,Yang2014,PhysRevA.101.013821}, is interesting in its own right, as spin noise reveals spectroscopic information about the atomic \cite{PhysRevApplied.17.L011001} or even the solid-state system under consideration \cite{PhysRevLett.95.216603,doi:10.1063/1.2794059,PhysRevLett.123.017401} in a non-perturbing way. Further, the study of spin noise in atomic vapors has elucidated quantum-non-demolition measurements \cite{KuzmichSpinSqueezingQND,shah-vasilakis,PhysRevLett.106.143601,MitchellNatureCom}, effects of atomic diffusion \cite{Diffusion,Ofer2020,PhysRevLett.96.043601}, noise studies at low-fields \cite{PhysRevA.104.063708,Sleator}, optical spin-noise amplification \cite{PhysRevApplied.9.034003}, SNR enhancement by squeezed-light \cite{Squeezed-Spin-Noise}, non-equilibrium SNS \cite{PhysRevLett.111.067201} and spin-alignment noise in a $^{133}$Cs vapor \cite{PhysRevResearch.2.012008}.

With respect to this work, however, the study of spontaneous spin noise in unpolarized vapors offers two additional advantages. First, it allows us to extrapolate the underlying physics to spin-projection noise in spin-polarized vapors \cite{Polzik}, and thus helps understand and define benchmarks against which any quantum sensing enhancements are compared with. Second, by using unpolarized vapors, in particular, a dual-species vapor, one can untangle various sorts of spin correlations, since they leave a clear signature in measurable spectral distributions of spin-noise power \cite{PhysRevA.106.023112}. 

Overlapping (dual-species) alkali-metal ensembles have been explored a while ago in the context of hybrid optical pumping \cite{ PhysRevLett.89.253002,PhysRevLett.105.243001}, specifically addressing the deterministic spin dynamics at non-zero spin-polarizations. More recently, two works \cite{Dellis,roy2015cross} studied spin-noise correlations and spin-noise transfer between two alkali species in an unpolarized vapor, arriving at conflicting results, while the authors in \cite{katz2015coherent} studied the strong SE-coupling of two alkali-metal species in a polarized vapor. Before summarizing the results of this work, we introduce some basic notions regarding atomic spin correlations.
\subsection{Atomic Correlations}
When it comes to correlations, one needs to define (I) the correlated degrees of freedom, (II) their dynamic aspect, (III) the various sub-types, and (IV) their quantum/classical character \cite{adesso2016measures}.
\begin{figure*}[t!]
\includegraphics[width=17 cm]{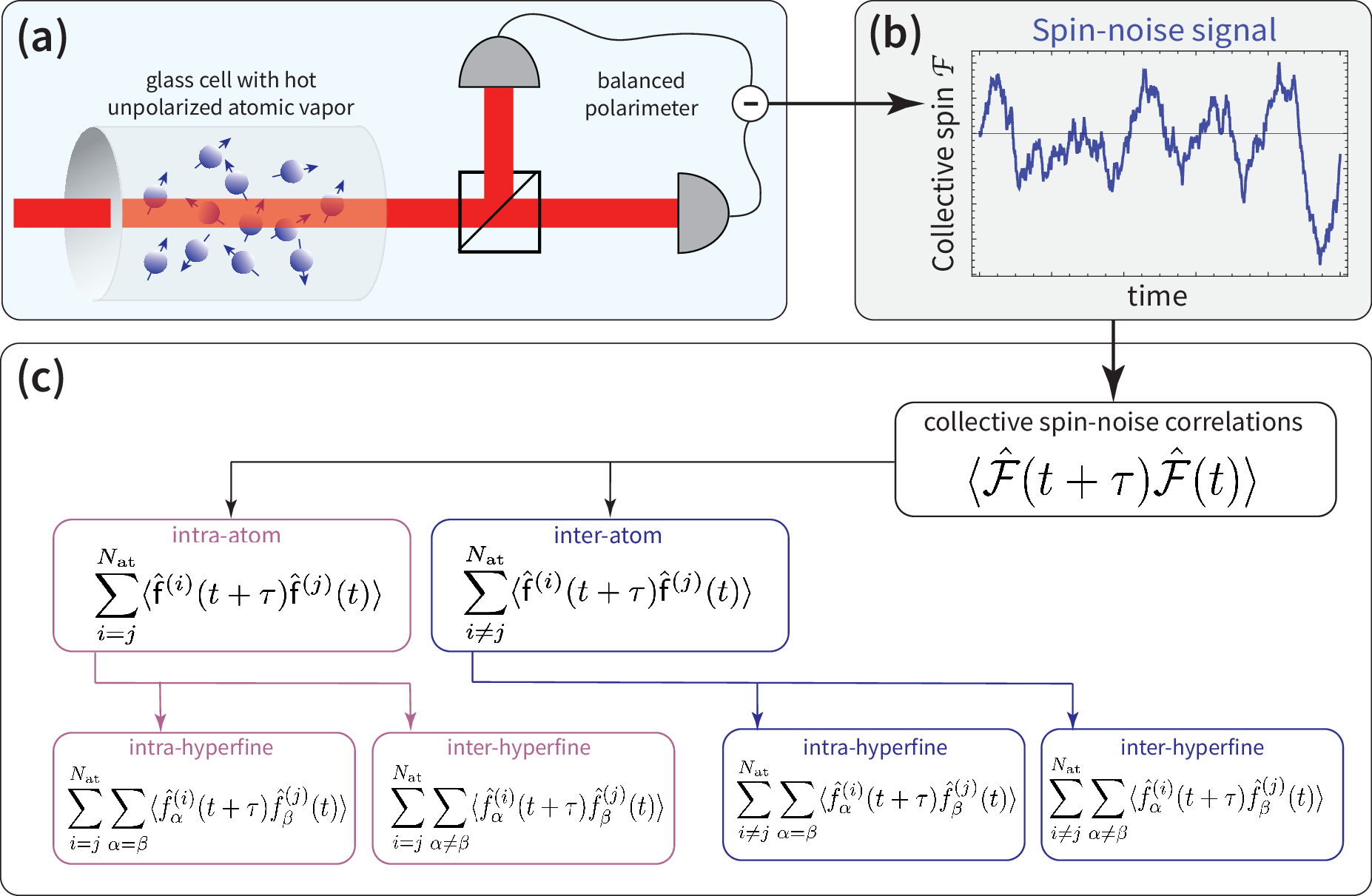}
\caption{Schematic diagram of atomic correlations measurable in a spin-noise experiment. (a) A hot and unpolarized alkali-metal vapor is non-destructively probed by a laser and a balanced polarimeter, comprised of a polarizing beam splitter and two photodetectors, recording the polarization fluctuations of the probe light. (b) The light's polarization fluctuations reflect the spontaneous fluctuations of the collective spin component, ${\cal F}$, along the probe beam direction. (c) The measurement directly leads to the correlator $\langle \hat{{\cal F}}(t+\tau)\hat{{\cal F}}(t)\rangle$, which contains an intra-atom part, $\sum_{i=j}^{N_{\rm at}}\langle \hat{\mathsf{f}}^{(i)}(t+\tau)\hat{\mathsf{f}}^{(j)}(t)\rangle$, and inter-atom part, $\sum_{i\neq j}^{N_{\rm at}}\langle \hat{\mathsf{f}}^{(i)}(t+\tau)\hat{\mathsf{f}}^{(j)}(t)\rangle$. Here $\hat{\mathsf{f}}^{(i)}$ derives from the probed component of the total spin operator of the $i$-th atom, and $N_{\rm at}$ the total number of atoms probed by the laser. In their turn, such correlations are further divided into intra- and inter-hyperfine correlations, where the indexes $\alpha,\beta$ take on the values $a=I+1/2$ and $b=I-1/2$, corresponding to the upper and lower hyperfine multiplet of an alkali-metal atom with nuclear spin $I$. Every correlator includes equal-time ($\tau=0$) and unequal-time ($\tau\neq 0$) correlations, each having a unique footprint on the spin-noise spectra and the spin-noise variances. In this work we add a further classification: inter-species versus intra-species correlations.}
\label{corrschem}
\end{figure*}
\subsubsection{Correlated degrees of freedom}
As we will show in detail in Section \ref{sec:theory}, what is experimentally accessible is the collective spin correlator 
\begin{equation}
\braket{\hat{{\cal F}}(t+\tau) \hat{{\cal F}}(t)}\label{corr}
\end{equation}
where $\hat{{\cal F}}(t)= \sum_{i=1}^{\NAtoms} \hat{{\mathsf f}}^{(i)} (t)$ is the collective spin of the ensemble along the direction of the laser beam, probing in total $\NAtoms$ atoms, with $\hat{{\mathsf f}}^{(i)} (t)$ being derived from the total spin of the $i$-th atom, $\mathbf{\hat{s}}+\mathbf{\hat{I}}$, where  $\mathbf{\hat{s}}$ is the electronic and  $\mathbf{\hat{I}}$ the nuclear spin.
\subsubsection{Dynamic aspect of correlations}
We  distinguish equal-time ($\tau=0$) from unequal-time correlations ($\tau\neq 0$), where $\tau$ is the time delay between the correlated observables. The former reflect total spin-noise power readily derived from the Wiener-Khinchin theorem, which relates the power spectral density of a signal with its correlation function. The total spin-noise power is obtained by the integral over frequencies of the power spectral density. 
\subsubsection{Sub-types of correlations}
The survival of correlations spontaneously building up by SE collisions in a hot rubidium vapor was studied in \cite{PhysRevA.103.L010401}, where it was theoretically shown that equal-time inter-atom correlations in specific states can be sustained amidst sequential SE collisions of the correlated partners with other atoms. 

In \cite{PhysRevA.106.023112} the authors studied unequal-time correlations generated by the dynamics of binary SE collisions at thermal equilibrium pertinent to spin-noise experiments with unpolarized vapors. The authors considered several kinds of correlations, schematically depicted in Fig. \ref{corrschem}. In particular, the correlator \eqref{corr} is composed of two terms $\sum_j\braket{\hat{\mathsf{f}}^{(j)}(t+\tau)\hat{\mathsf{f}}^{(j)}(t)}$ and $\sum_{i\neq j}\braket{\hat{\mathsf{f}}^{(i)}(t+\tau)\hat{\mathsf{f}}^{(j)}(t)}$, describing intra- and inter-atom correlations, respectively. For each such kind one can further  distinguish  intra-hyperfine from inter-hyperfine correlations. To describe those, the total single-atom spin operators are written as $\hat{f}_{\alpha}^{(i)}(t)$, where  $\hat{f}_{\alpha}$ denotes the projection of the total atom's spin, $\mathbf{\hat{s}}+\mathbf{\hat{I}}$, onto the hyperfine manifold $\alpha$, with $\alpha \in \{a,b\}$ denoting either the upper ($a \equiv I+1/2$) or the lower ($b \equiv I-1/2$) ground-state hyperfine manifold. Thus, as shown in Fig. \ref{corrschem}, we resolve the original correlator in four kinds of correlations: (a) intra-atom and intra- or inter-hyperfine correlations, and (b) inter-atom and intra- or inter-hyperfine correlations. 

In this work we bring into the picture an additional type, the inter-species correlations. In a single-species vapor the inter-atom correlations always contribute to the spin-noise spectrum simultaneously with the intra-atom correlations, and the two are hardly distinguishable. {\it Here comes the usefulness of a dual-species vapor, in which the inter-species inter-atom correlations can be experimentally distinguished and thus shed light on inter-atom correlations even for the single-species case}.
\subsubsection{Character of correlations}
The quantum/classical character of the correlations is a subtle issue which appears to have a parametric dependence, the parameter being the vapor's spin-polarization. An example, drawn from quantum information science, demonstrating a parametric dependence of correlations is the two-qubit Werner state $\rho=\frac{1}{4}(\mathbb{1}-\alpha\boldsymbol{\hat{\sigma}}\otimes\boldsymbol{\hat{\sigma}})$, where $\boldsymbol{\hat{\sigma}}\otimes\boldsymbol{\hat{\sigma}}=\hat{\sigma}_x\otimes\hat{\sigma}_x+\hat{\sigma}_y\otimes\hat{\sigma}_y+\hat{\sigma}_z\otimes \hat{\sigma}_z$, $\mathbb{1}$ is the identity and $\hat{\sigma}_i$, $i \in \{x,y,z\}$ are the Pauli operators \cite{Popescu,Krammer}. Since the term $\boldsymbol{\hat{\sigma}}\otimes\boldsymbol{\hat{\sigma}}$ is traceless, the spin populations of both qubits in each of the two states along the quantization axis (e.g. the states with $\sigma_z=\pm 1$) are 1/2. Thus the average spin is zero. Yet the state $\rho$ exhibits correlations for any $\alpha \neq 0$, since $\braket{\hat{\sigma}_z\otimes \hat{\sigma}_z}=-\alpha$. In particular, the correlations are quantum and violate the CHSH inequality for $\alpha >1/\sqrt{2}$. Thus, the character of the correlations of this 2-qubit state depends on the parameter $\alpha$. In Sec. \ref{sec:ent} we will further comment on this point in regard to this work.
\subsection{Summary of the results of this work}
In this work we undertake (A) an experimental and (B) a formal theoretical study of spin-noise correlations in a \textsuperscript{87}\rm{Rb}-\textsuperscript{133}\rm{Cs}  hot vapor. (A1) We unambiguously demonstrate the existence of spontaneously generated unequal-time inter-species spin-noise correlations, driven by spin-exchange collisions. (A2) Under certain conditions, i.e. finite measurement bandwidth, we also demonstrate the existence of inter-species correlations reflected in a total spin-noise power different from what would be obtained were the atoms uncorrelated. (B1) Using the full density matrix description of spin-exchange and spin-relaxation dynamics, we develop a first-principles theoretical framework that captures the subtle physics of all four sub-types of correlations and resolves inconsistencies of previous works \cite{Dellis,roy2015cross}. (B2) We draw qualitative conclusions about the character of inter-species spin-noise correlations in spin-polarized dual-species vapors.
 \begin{figure*}[ht]
\includegraphics[width=17 cm]{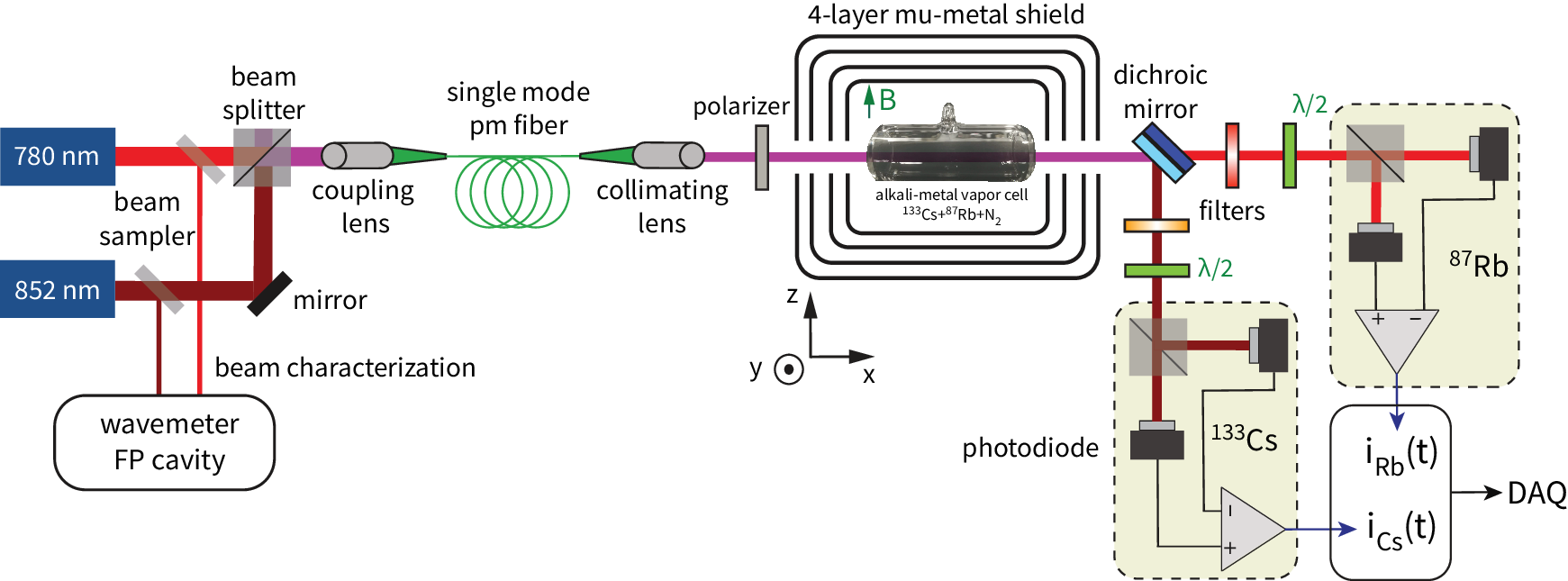}
\caption{Experimental setup for the measurement of \textsuperscript{87}\rm{Rb}-\textsuperscript{133}\rm{Cs} spin-noise correlations. Two single-mode external cavity diode lasers, blue-detuned from the corresponding atomic resonances, are combined in a single-mode optical fiber and directed towards the vapor cell, held inside a 4-layer mu-metal magnetic shield. A dichroic mirror after the cell is used to spatially separate the two overlapping wavelengths exiting the atomic medium, while optical filters further prevent leakage of the unwanted light to the detectors. Two identical balanced polarimeters comprised of a half-waveplate, a polarizing beam splitter cube, and a balanced photodetector provide the signals feeding the data acquisition system. Both wavelengths were monitored within $10$ \si{\mega\hertz} resolution using a Fizeau wavelength-meter (High Finesse WS7). Single-mode operation was additionally monitored by a second scanning Fabry-Perot interferometer.}
\label{setup}
\end{figure*} 
\section{\label{sec:exp}Experimental setup, raw data and observables}
In this section we describe the experimental setup, and define the experimental observables to be used in the theoretical treatment of the following section.

\subsection{Experimental setup}
The experimental setup is shown in Fig. \ref{setup}. A cylindrical cell of diameter \SI{12.7}{\milli \meter} and length \SI{47}{\milli \meter}, with anti-reflection coated windows, contains 330 Torr of N\textsubscript{2} buffer gas and a metallic droplet of \textsuperscript{133}Cs and \textsuperscript{87}Rb in molar ratio 28:72. The cell resides in a ceramic oven heated with \SI{200}{\kilo\hertz} current to \SI{160}{\celsius}, resulting in number densities $n_{_{\mathrm{Cs}}} \approx n_{_{\mathrm{Rb}}}\approx \SI{e14}{\per\centi\meter\cubed}$.

In the presence of nitrogen the optical linewidths are pressure-broadened (FWHM) to $7.34$ \si{\giga\hertz} for $^{133}$Cs and $7.08$ \si{\giga\hertz} for $^{87}$Rb, therefore the two ground-state hyperfine levels are moderately resolved. The collective atomic spins of the alkali-metal species are probed by two single-mode external cavity diode lasers (Toptica DL-780 pro and DL-850 pro) with their wavelengths blue-detuned from the $\rm{D}_2$ transition by several tens of \si{\giga\hertz}. Both linearly polarized laser beams enter in the same single-mode optical fiber producing a two-color  Gaussian beam with  diameter $\sim 2$~mm ($1/e^2$ intensity width) at the position of the cell (with a small difference in the two colors of less than $10 \%$ due to diffraction).

The vapor cell is placed inside a $4$-layer mu-metal shield, protecting the atomic spins from ambient magnetic fields. Within the shields, a coil system generates a DC magnetic field $\mathbf{B}=(0,0,B)$ transverse to the beam-propagation axis, ranging from $4$ mG to $92$ mG, with the corresponding spin precession frequencies reaching several tens of \si{\kilo\hertz}.  

At the exit of the cell the two-color beam is incident on a dichroic mirror, which directs each color to a separate balanced polarimeter, the two polarimeters being otherwise identical. The use of 780~\si{\nano\meter} and 852~\si{\nano\meter} filters further reduces light leakage from one wavelength to the other, resulting in negligible cross-talk in the polarization detection of the two beams.  Each balanced photodetector has detection efficiency $\eta$.

The cesium and rubidium polarimeter signals are acquired and analyzed in the frequency domain by feeding them into a spectrum analyzer (Stanford Research Systems SR780 Dynamic Signal Analyzer), which provides the cross-correlation power spectral density and the two single-channel auto-correlation power spectra. The frequency resolution of the FFT is $800$ bins, with a sampling rate of $262$ \si{\kilo\hertz}. A Hann window is applied and an analog anti-aliasing filter applied before digitization eliminates all frequency components above $102.4$ \si{\kilo\hertz}. 
\subsection{\label{sec:signals_a} Auto-correlation and cross-correlation observables}
Each balanced photodetector receives a continuous optical signal and generates a corresponding photo-current, which we denote by $i_{\beta} (t)$, with $\beta \in \{\Rb,\Cs\}$. Spin noise (SN) and photon-shot-noise render $i_{\beta} (t)$ a classical and real stochastic process. We record the signals for long enough time such that both spin-noise and photon-shot-noise have stationary statistical moments. We are interested in the cross-correlation between the two balanced photodetector signals:
\begin{equation}
C_{\Rb,\Cs}(\tau) \equiv \langle \langle i_{\Rb}(t+\tau) i_{\Cs} (t)  \rangle \rangle,
\end{equation} 
where $\langle \langle \cdot \rangle \rangle$ denotes the average over all possible realizations of the stochastic measurement outcomes. Stationarity implies that the correlation does not depend on the absolute time $t$. 

The quantum-mechanical operator associated with the measured photocurrent is:
\begin{equation}
\begin{split}
\hat{I}_{\beta}(t) &= q_e \eta_{\beta} \left( \hat{a}_{k,+45}^{\dag} \hat{a}_{k,+45}- \hat{a}_{k,-45}^{\dag} \hat{a}_{k,-45}\right)_{\beta}\\
&= 2 q_e \eta_{\beta} \hat{{\cal S}}_{2,\beta} (t),
\end{split}
\end{equation}
where $\hat{a}_{k,\ell}\equiv \hat{a}_{k,\ell} (t)$ and $\hat{a}^{\dag}_{k,\ell}\equiv \hat{a}^{\dag}_{k,\ell} (t)$ are respectively the travelling-wave annihilation and creation operators for the mode $k$ and polarization $\ell$, normalized so that $\hat{a}^{\dag}_{k,\ell} \hat{a}_{k,\ell}$ is the photon flux  in the respective mode \cite{PhysRevA.89.033850}. The index $k$ labels the longitudinal as well as the spatial mode, the latter being defined by the spatial mode of the probe beam (Gaussian $\text{TEM}_{00}$ in the experiment). The $\pm45^\circ$ linear polarizations are characterized with respect to the polarization direction of the probe beam before the interaction with the atomic medium, chosen to be along the $\mathbf{\hat{y}}$-direction. Finally, $\hat{{\cal S}}_2 (t)$ is the optical Stokes operator given by:
\begin{align}
\hat{{\cal S}}_2 (t)&=  \frac{ \hat{a}_{k,+45}^{\dag} \hat{a}_{k,+45}- \hat{a}_{k,-45}^{\dag} \hat{a}_{k,-45} }{2}\nonumber\\
&=  \frac{ \hat{a}_{k,y}^{\dag} \hat{a}_{k,z}+\hat{a}_{k,z}^{\dag} \hat{a}_{k,y} }{2},
\end{align}
where the coordinate system is defined in Fig.~\ref{setup}. The operators $\hat{{\cal S}}_{2,\beta}$ refer to the Stokes operator $\hat{{\cal S}}_2$ for each of the two wavelenghts probing atom species $\beta$.

The cross-correlation can be therefore expressed as \cite{gardiner2004quantum, Glauber63}:
\begin{equation}
\begin{split}
C_{\Rb,\Cs}(\tau) & = \langle : \hat{I}_{\Rb}(t+\tau) \hat{I}_{\Cs}(t) : \rangle \\ 
&= 4 q_e^2 \eta_{\Rb} \eta_{\Cs} \langle : \hat{{\cal S}}_{2, \Rb}(t+\tau) \hat{{\cal S}}_{2, \Cs}(t): \rangle,
\label{eq:Xcorr1}
\end{split}
\end{equation}
where $::$ denotes normal ordering. In the above equation, the quantum operators are written in the Heisenberg picture and the single brackets $\langle \cdot \rangle$ denote quantum mechanical  expectation value.
Taking into account the bosonic commutation relations \cite{gardiner2004quantum, Glauber63}, and considering that the off-resonance interaction with the atomic vapor is to a very good approximation a linear optical process and thus does not mix annihilation and creation operators, we can make use of the unequal-time commutation relations:
\begin{equation}
\left[ \hat{a}^{\dag}_{k,\ell}(t),\hat{a}^{\dag}_{k',\ell'}(t')\right]=\left[ \hat{a}_{k,\ell}(t),\hat{a}_{k',\ell'}(t')\right] =0,
\end{equation}
and show $\langle : \hat{I}_{\Rb}(t+\tau) \hat{I}_{\Cs}(t) : \rangle= \langle : \hat{I}_{\Cs}(t) \hat{I}_{\Rb}(t+\tau)  : \rangle$. In order to emphasize that the correlation function is invariant under the preceding exchange of operators we express Eq.\eqref{eq:Xcorr1} in the symmetrized form \cite{Glauber63}:
\begin{align}
C_{\Rb,\Cs}(\tau) &= 2 q_e^2 \eta_{\Rb} \eta_{\Cs}\Big[ \langle  \hat{{\cal S}}_{2, \Rb}(t+\tau) \hat{{\cal S}}_{2, \Cs}(t) \rangle \nonumber \\
&+ \langle \hat{{\cal S}}_{2, \Cs}(t)  \hat{{\cal S}}_{2, \Rb}(t+\tau) \rangle \Big].
\label{eq:cross-corr}
\end{align}
The cross-spectral density function is the Fourier transform of the cross-correlation:
\begin{equation}
\tilde{C}_{\Rb,\Cs} (\nu) = \int_{-\infty}^{\infty} C_{\Rb,\Cs}(\tau) e^{-i 2\pi \nu \tau} d\, \tau.\label{Cnu}
\end{equation}
The spectrum analyzer estimates the power spectral density (PSD) by performing a finite Fourier transform for a specified record time $T$ and averages the product of the Fourier components over the different realizations of the stochastic processes \cite{SR780}:
\begin{align}
& \tilde{C}^{\text{SA}}_{\Rb,\Cs} (\nu) = \nonumber\\ 
& \frac{1}{T} \langle \langle \left(  \int_0^T  i_{\Rb}(t)  e^{-i 2 \pi \nu t} \, d t  \right)^{*} \int_0^T i_{\Cs}(s) e^{-i 2 \pi \nu s}    \, d s\rangle \rangle \nonumber\\ 
& = \int_{-T}^{T} C_{\Rb,\Cs}(\tau) e^{-i 2 \pi \nu \tau}\left( 1-\frac{|\tau|}{T} \, d\tau\right).\label{eq:psd_SA}
\end{align}
To derive the second line, the variables of integration were changed from $(t,s)$ to $(t,\tau=t-s)$ and the region of integration was modified accordingly \cite{engelberg2018random}. Thus, the spectrum analyzer performs an exact evaluation of the spectral density (Eq. \eqref{Cnu}) in the limit of infinite long sample realization $T\rightarrow + \infty$. In practice, for an accurate approximation, it is sufficient to choose the sample length to be much larger than the time difference $\tau$ at which the correlation function becomes negligible. 

Similarly, single-channel auto-correlation functions and their corresponding power spectral densities are defined as $C_{\beta,\beta} (\tau) = \langle \langle i_{\beta}(t) i_{\beta}(t+\tau) \rangle \rangle$ and $\tilde{C}_{\beta,\beta} (\nu) = \int_{-\infty}^{\infty} C_{\beta,\beta} (\tau) e^{-i 2 \pi \nu \tau} \, d \tau $, respectively, with $\beta \in\{\Rb,\Cs\}$. As before, the auto-correlation function can be expressed in the symmetrized form:
\begin{align}
C_{\beta,\beta} (\tau) &= 2 q_e^2 \eta_{\beta} \times \nonumber\\
& \left( \langle \hat{{\cal S}}_{2,\beta}(t)  \hat{{\cal S}}_{2,\beta}(t+\tau) \rangle+ \langle \hat{{\cal S}}_{2,\beta}(t+\tau)  \hat{{\cal S}}_{2,\beta}(t) \rangle \right).\label{Cbetabeta}
\end{align}
\section{\label{sec:theory} Theoretical Derivation of Cross- and Auto-Correlation Spectra}
The theoretical description of the cross correlations is based on treating the atom-light coupling, and the atomic state evolution due to the atom's hyperfine structure hamiltonian, atom-atom spin-exchange collisions, and a number of additional spin relaxation phenomena. Such topics have been described in detail elsewhere \cite{PhysRev.163.12,stockton,Deutsch2010,PhysRevA.77.032316}, and we recapitulate the description for the sake of completeness in Appendix \ref{sec:AppA}. In the following two subsections we present the results of extending such treatments to the case of the two-species atomic vapor treated herein.
\subsection{Atom-light coupling}
Based on the light-atom interaction dynamics, it follows that for probe light linearly polarized in the $y$-direction, the solution of the Heisenberg equation of motion for the Stokes operator yields in the small-angle approximation\cite{PhysRevA.71.032348,Kuzmich}
\begin{equation}
\hat{{\cal S}}_2^{(\text{out})} (t) \approx \hat{{\cal S}}_2^{(\text{in})} (t)+\frac{1}{2}\Phi \bar{g} \mathcal{\hat{F}} (t)\label{eq:S2out},
\end{equation} 
where $\Phi$ is the photon-flux (photons per unit time). The out (in) superscripts denote the operator after (before) interacting with the atomic sample, $\bar{g}= (|g_a|+|g_b|)/2$ is the mean coupling constant to the two ground-state hyperfine manifolds and $\mathcal{\hat{F}} (t)=\sum_{i=1}^{\NAtoms}\hat{\mathsf{f}}^{(i)}_x(t)$ is the probed collective atomic spin with
\begin{equation}
\hat{\mathsf{f}}^{(i)}_x(t)= \frac{g_a}{\bar{g}} \hat{f}_{a,x}^{(i)} (t) - \frac{g_b}{\bar{g}} \hat{f}_{b,x}^{(i)} (t)
\label{collective_F} 
\end{equation}
Here $\hat{f}_{a,x}^{(i)}$ and $\hat{f}_{b,x}^{(i)}$ are the projections of the $x$-component of the total spin of the $i$-th atom, $\hat{s}_x+\hat{I}_x$, onto the upper ($a=I+1/2$) and lower ($b=I-1/2$) hyperfine multiplet, respectively, with the corresponding coupling constants being $g_a$ and $g_b$ (see Appendix \ref{sec:AppA}, where we also generalize Eq. \eqref{collective_F} to include a non-uniform intensity distribution of the light). 

The ${\rm in}\rightarrow {\rm out}$ change of $\hat{{\cal S}}_2$ described by Eq. \eqref{eq:S2out} reflects polarization rotation of light (Faraday rotation) at an angle $\phi_{\text{FR}} = \frac{1}{2} \bar{g} \braket{\mathcal{\hat F}}$. Thus, Eq. \eqref{eq:S2out} connects the Stokes operator ${\cal S}_2(t)$ to the underlying spin operator $\mathcal{\hat{F}}(t)$. Using Eq. \eqref{eq:cross-corr}, Eq. \eqref{Cbetabeta}, and Eq. \eqref{collective_F}, we arrive at the following expressions connecting the measured correlations (cross-correlation and auto-correlations) to the underlying collective spin correlators:
\begin{align}
C_{\Rb,\Cs}(\tau) & = \frac{q_e^2 \eta_{\Rb} \eta_{\Cs} \Phi_{\Rb} \Phi_{\Cs} \bar{g}_{\Rb} \bar{g}_{\Cs}}{2} \times \nonumber\\
 & \langle \mathcal{\hat{F}}_{\Rb}(t+\tau) \mathcal{\hat{F}}_{\Cs}(t)+ \mathcal{\hat{F}}_{\Cs}(t) \mathcal{\hat{F}}_{\Rb}(t+\tau) \rangle,
 \label{eq:cross-species corr}\\
 C_{\beta,\beta}(\tau)&= 2 q_e^2 \eta_{\beta} \Phi_\beta \Big[ \delta(\tau)\nonumber\\
 &+ \frac{1}{2} \eta_{\beta} \Phi_\beta \bar{g}^2_{\beta} \langle \mathcal{\hat{F}}_{\beta}(t+\tau)  \mathcal{\hat{F}}_{\beta}(t) \rangle  \Big], \label{eq:AutoCorrelationPolarimetry}
\end{align}
where $\beta \in \{\Rb,\Cs \}$. We note that we have taken into account that the polarization properties of the two beams before the interaction with the atoms are uncorrelated, i.e. $\langle \hat{{\cal S}}^{(\rm{in})}_{2,\Rb}(t) \hat{{\cal S}}^{(\rm{in})}_{2,\Cs}(t')\rangle =0$.
\subsection{Dynamics of mean spin}
\label{sec:DynamicsMeanSpin}
By considering the various processes that affect the atomic spin evolution (see Appendix \ref{sec:MeanSpinDynamics}), we arrive at two coupled density matrix evolution equations for Rb and Cs:
\begin{align}
\frac{d }{dt} \rho_\Rb &= A_\Rb \hat{\mathbf{I}}_\Rb \cdot \hat{\mathbf{s}}_\Rb+g_s \mu_B \hat{\mathbf{s}}_\Rb \cdot \mathbf{B}+R \left( \phi_\Rb-\rho_\Rb\right)\nonumber\\ 
&+R_{\rm se}^{\Rb,\Rb} \left\{ \phi_\Rb \left (1+4 \langle \hat{\mathbf{s}}_\Rb \rangle \cdot \hat{\mathbf{s}}_\Rb \right ) - \rho_\Rb \right \}\nonumber\\
&+R_{\rm se}^{\Rb,\Cs} \left\{ \phi_\Rb \left (1+4 \langle \hat{\mathbf{s}}_\Cs \rangle \cdot \hat{\mathbf{s}}_\Rb \right ) - \rho_\Rb \right \}
\label{drdtRb}\\
\frac{d }{dt} \rho_\Cs &= A_\Cs \hat{\mathbf{I}}_\Cs \cdot \hat{\mathbf{s}}_\Cs+g_s \mu_B \hat{\mathbf{s}}_\Cs \cdot \mathbf{B}+R \left( \phi_\Cs-\rho_\Cs\right)\nonumber\\ 
&+R_{\rm se}^{\Cs,\Cs} \left\{ \phi_\Cs \left (1+4 \langle \hat{\mathbf{s}}_\Cs \rangle \cdot \hat{\mathbf{s}}_\Cs \right ) - \rho_\Cs \right \}\nonumber\\
&+R_{\rm se}^{\Cs,\Rb} \left\{ \phi_\Cs \left (1+4 \langle \hat{\mathbf{s}}_\Rb \rangle \cdot \hat{\mathbf{s}}_\Cs \right ) - \rho_\Cs \right \}
\label{drdtCs}
\end{align} 
where $A_\beta$ is the hyperfine coupling, the rate $R$ includes all the relaxation processes, other than the spin-exchange relaxation, that destroy electron polarization without affecting the nucleus, $\phi_\beta$ is the atom's density matrix with zero electronic polarization, and $R_{\rm se}^{\beta,\gamma}$ is the spin-exchange rate transferring spin-polarization from atomic species $\gamma$ to species $\beta$, with $\beta,\gamma\in \{\Rb,\Cs\}$. The diffusion of atoms out of the probe beam must also be included in these dynamics (see Appendix \ref{sec:MeanSpinDynamics}).

Multiplying both sides of Eqs. \eqref{drdtRb} and \eqref{drdtCs} by the spin operator $\hat{f}_{i}$ (here $i$ is a general index for identifying the atomic species, the hyperfine manifold and the Cartesian component) and taking the trace, the dynamics of $\langle \hat{f}_i \rangle$ are determined and  a closed system of equations can be derived for the time evolution $\langle \hat{f}_i \rangle$. Like in \cite{PhysRevA.16.1877, katz2015coherent, PhysRevA.106.023112}, we define the vector $\mathbf{\hat{X}}(t)$ for the transverse collective-spin components of each species and each hyperfine level:
\begin{equation}
\mathbf{\hat{X}} (t) \equiv \left[ \hat{f}_{a,x}^\Rb,\hat{f}_{a,y}^\Rb,\hat{f}_{b,x}^\Rb,\hat{f}_{b,y}^\Rb, \hat{f}_{a,x}^\Cs,\hat{f}_{a,y}^\Cs,\hat{f}_{b,x}^\Cs,\hat{f}_{b,y}^\Cs  \right]^{\intercal},
\end{equation}
where $\hat{f}_{\alpha,q}^{\beta}$ refers to the total atomic spin of species $\beta\in \{\Rb,\Cs\}$ along the $q$-axis with $q\in\{x,y\}$, and in the hyperfine state $\alpha\in\{a,b\}$. 
The density matrix evolution Eqs. \eqref{drdtRb} and \eqref{drdtCs} contain nonlinear terms proportional to $\langle\hat{\mathbf{s}}_\beta \rangle \cdot \hat{\mathbf{s}}_\gamma$ associated with the SE interaction; however, for noise measurements around zero mean spin-polarization we linearize such terms by keeping only first-order contributions from the fluctuations. This approximation leads to the linear evolution equation 
\begin{equation}
\frac{d}{dt} \langle \mathbf{\hat{X}}(t) \rangle = A \langle \mathbf{\hat{X}} (t)\rangle, \label{MatrixEquationMeanValues1}
\end{equation}  
where the drift matrix $A$ is comprehensively derived in Appendices \ref{sec:MeanSpinDynamics} and \ref{sec:AppC}. 
\subsection{Spin correlations in the time domain}
The estimation of the spin-noise spectrum requires the evaluation of correlators $\langle\hat{f}_i(t+\tau) \hat{f}_j(t) \rangle$. To find those, we use the quantum regression theorem \cite{Carmichael2008, gardiner2004quantum}, which states that if the expectation values of a set of observables $\hat{M}_\mu$, $\mu=1,2,...$ follow a coupled set of linear equations: 
\begin{equation}
\frac{d}{dt} \langle \hat{M}_\mu(t) \rangle = \sum_{\lambda} A_{\mu \lambda} \langle \hat{M}_{\lambda}(t)\rangle, \label{eq:QRTMeanEvol}
\end{equation}
then the two-time correlation functions satisfy the equations (for $\tau \geq 0$):
\begin{align}
\frac{d}{d \tau} \langle \hat{M}_{\kappa}(t) \hat{M}_\mu (t+\tau) \rangle &= \sum_\lambda A_{\mu \lambda} \langle \hat{M}_{\kappa}(t) \hat{M}_\lambda (t+\tau) \rangle, \\
\frac{d}{d \tau} \langle \hat{M}_\mu (t+\tau) \hat{M}_{\kappa}(t)  \rangle &= \sum_\lambda A_{\mu \lambda} \langle \hat{M}_\lambda (t+\tau) \hat{M}_{\kappa}(t)  \rangle.
\end{align}
The regression theorem holds when the equations of motion for the expectation values are linear and the system-environment correlations can be neglected \cite{Swain1981, Dmcke1983}. The justification for linearity arises from considering small fluctuations as noted previously. For the second requirement, we argue that the environment, where spin-information is lost, is associated with the (abstract) space spanned with all the collisional parameters. As long as a large number of particles are probed, when the ensemble average is taken over all the different types of collisions, the correlations between the collective spin-system and the environment are lost, thus rendering the regression theorem applicable.

We thus arrive at the symmetrized and real covariance matrix: 
\begin{equation}
\mathcal{R} (\tau) = \frac{1}{2} \Big( \langle \mathbf{\hat{X}}(t) \mathbf{\hat{X}}^{\intercal}(t+\tau) \rangle + [\langle \mathbf{\hat{X}}(t+\tau) \mathbf{\hat{X}}^{\intercal}(t) \rangle ]^{T} \Big).
\end{equation}
The quantum regression approach yields for $\tau \geq 0 $: 
\begin{equation} 
\frac{d}{d \tau} \mathcal{R} (\tau) = A \mathcal{R}(\tau) \rightarrow \mathcal{R}(\tau) = e^{A \tau} \mathcal{R}(0). \label{eq:RtEqExpAtR0}
\end{equation}
Given that $\mathcal{R}(0)$ is symmetric, for $\tau<0$ it is $\mathcal{R}(\tau)= \mathcal{R}(-\tau)^{T}$. 

In the following we extend the analysis of \cite{PhysRevA.106.023112} to the current case of a dual-species vapor. The matrix $A$ is diagonalizable, so we can write $e^{A \tau} =  V e^{\Lambda \tau} V^{-1}$, where $V$ is the matrix whose $i$-th column is the $i$-th eigenvector of $A$, and $\Lambda$ is a diagonal matrix with its diagonal elements being the corresponding eigenvalues of $A$. Therefore, according to Eq.\eqref{eq:RtEqExpAtR0}, the time dependence for any spin correlation of the form $\langle \hat{f}_{\alpha,q}^{\beta} (0) \hat{f}_{\alpha',q'}^{\beta'} (0)\rangle $ manifests as a summation over all the eigenvalues of $A$ of exponentials of the form $e^{\lambda_k \tau}$, where $\lambda_k$ is the $k$-th eigenvalue of $A$. The eigenvalues of $A$ are in general complex numbers, but since $A$ has real elements, the eigenvalues appear in complex conjugate pairs. Since $A$ is an $8 \times 8$ matrix, there are 4 pairs of eigenvalues, hence the elements of the covariance matrix can be written (for $\tau \geq 0$): 
\begin{equation}
\mathcal{R}^{\beta,\beta'}_{\alpha q,\alpha' q'}(\tau)=\sum_{k=1}^4 \Re \left[ c_k(\beta\alpha q;\beta'\alpha' q') e^{-\Gamma_k \tau+i \Omega_{k} \tau} \right], \label{eq:CovarianceMatrixComponentsSum}
\end{equation}
where $\Gamma_k =\Re \left[ \lambda_k \right]+R_D$ and $\Omega_{k}=\Im \left[ \lambda_k \right]$ are the decoherence rate and the precession frequency, respectively, of mode $k$. In the rates $\Gamma_k$ we took into account the relaxation rate, $R_D$, due to atoms diffusing out of the probe beam. The coefficients $c_k$ are complex numbers that depend on the steady-state ($\tau=0$) covariance $R(0)$. 
\subsection{Spin correlations in the frequency domain}
The spectrum, obtained by the Fourier transform of the correlation function (see Appendix \ref{sec:AppC}), is then given as a sum of complex Lorentzians or equivalently as the sum of dispersive and Lorentzian functions:
\begin{align}
S^{\beta,\beta'}_{\alpha q,\alpha' q'}(\nu) = \sum_{k=1}^4  &\zeta_k(\beta\alpha q;\beta'\alpha' q') \frac{\Gamma_k}{(\nu-\nu_{k})^2+(\Gamma_k/2 \pi)^2}+ \nonumber \\
 & \zeta_k'(\beta\alpha q;\beta'\alpha' q') \frac{\nu-\nu_k}{(\nu-\nu_{k})^2+(\Gamma_k/2 \pi)^2}, \label{eq:CrossCorrelationSpectrum0}
\end{align}
where $\nu_k=\Omega_k/2 \pi$. In contrast to the auto-correlation spectrum, the cross-correlation spectrum of Eq.\eqref{eq:CrossCorrelationSpectrum0} is in general complex, i.e. the coefficients $\zeta_k$, $\zeta_k'$ can be complex for $\beta \neq \beta'$. Therefore, in order to obtain all the information related to the spectrum, both the real and imaginary components should be recorded. 

Finally we note that the covariance matrix $\mathcal{R}(0)$ entering Eq.\eqref{eq:RtEqExpAtR0} can be evaluated by integrating the spectrum over all frequencies (see Appendices \ref{sec:AppA} and \ref{sec:app:BandwidthEffect}). In practice however, the measurement is performed over a limited frequency range either due to the finite-bandwidth of the electronics/detectors or because deliberately the experimental application requires a limited sampling rate. Depending on the experimental conditions, this may affect the apparent correlations as discussed in Sec. \ref{sec:results}.
\section{\label{sec:exp_analysis}Experimental data with theoretical fits}
We acquire $^{87}$Rb-$^{133}$Cs cross-correlation spectra, as well as single-species $^{87}$Rb and $^{133}$Cs power spectra, for six different magnetic fields: 4~mG, 6~mG, 12~mG, 24~mG, 46~mG and 92~mG. We verify that the sign of the cross-correlation signal makes physical sense. Indeed, the experimental observable (balanced polarimeter output) is a product of the measured atomic spin with an atom-light coupling factor. The latter has a sign depending on the probe detuning from the atomic resonance (see Appendix A). The data are acquired with both probe beams being blue detuned from the corresponding atomic resonance. As a result, the atom-light scaling factors do not alter the overall sign of the cross-correlation. Nevertheless, we do verify that by flipping either one wavelength detuning from blue to red, the cross-spectra change sign. For the accurate interpretation of the positivity or negativity of the dual-species spin correlations, it is also necessary to measure the direction of optical rotation consistently for both beams. We do this by introducing an achromatic waveplate in the common path of the two beams, confirming that the two polarimeter outputs change in the same way with the rotation of the waveplate.
\begin{figure*}[th]
\includegraphics[width=17.5cm]{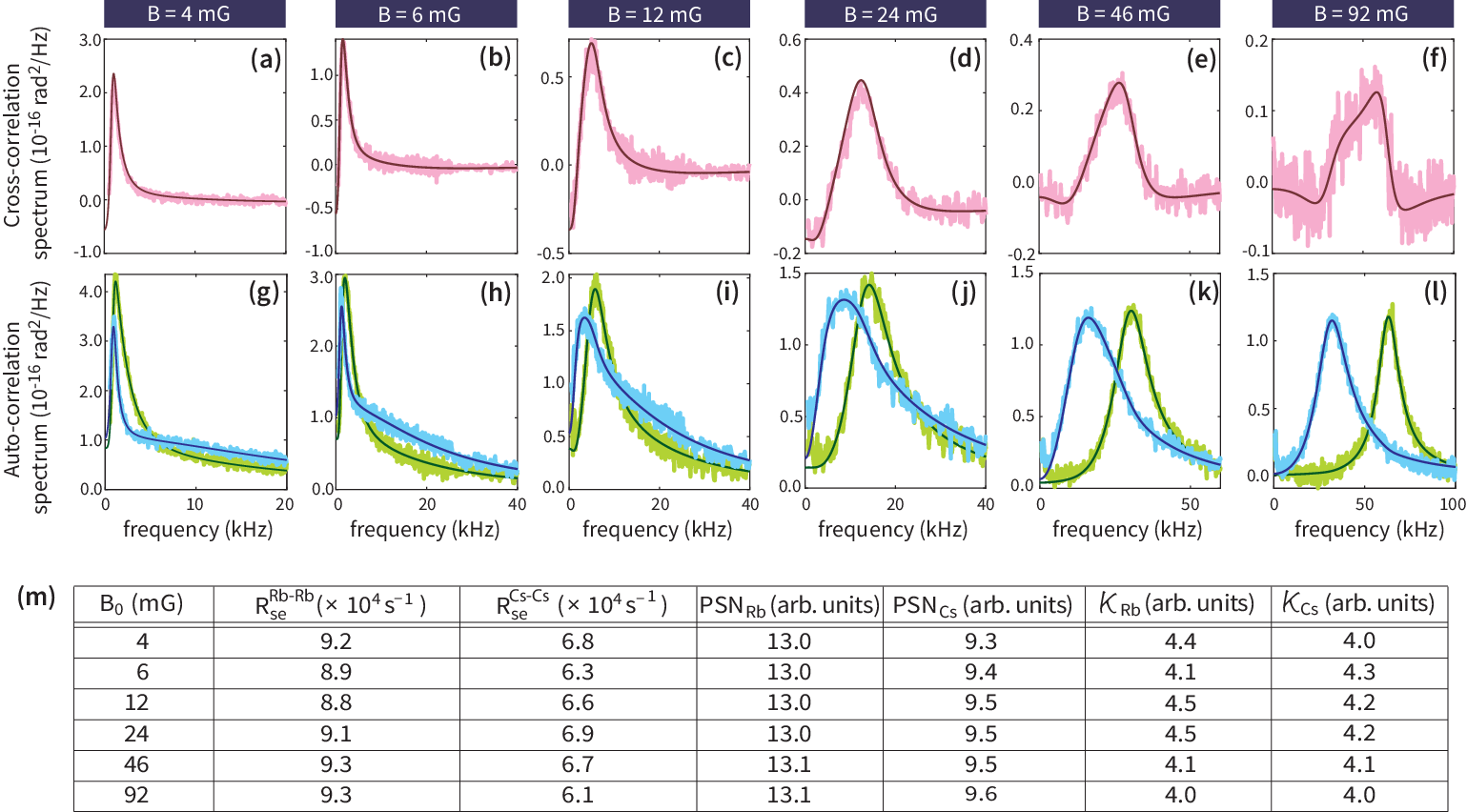}
\caption{Measured cross correlation spectra (a)-(f) and auto-correlation power spectra (g)-(l) for the \textsuperscript{87}Rb (green, higher frequencies) and \textsuperscript{133}Cs (cyan, lower frequencies) spin-ensembles at six different magnetic fields, together with theoretical fits (solid lines). Each spectrum is the average of 5000 runs. With increasing magnetic field, the magnitude of the cross-correlation peak is seen to drop, since when the difference in precession frequencies of the two species is larger than the magnetic linewidth, the spin-exchange coupling of the two different states is averaged out. The narrowing of the auto-correlation spectra is also evident, as by reducing the magnetic field spin dynamics gradually enter the SERF regime. (m) Collection of fitting parameters for magnetic field, self spin-exchange rates, photon shot noise levels and scaling factors. The magnetic field aside, all parameters are consistent across the six magnetic field values.}
\label{fig:CrossCorrelationData}
\end{figure*}

To fit the theoretical model to the data we use as model spectrum the expression
\begin{align}
 S_{\text{model}}^{\Rb, \Cs}(\nu)  \propto  & g_{a}^{\Rb} g_{a}^{\Cs} S_{ax,ax}^{\Rb,\Cs}(\nu) - g_{a}^{\Rb} g_{b}^{\Cs} S_{ax,bx}^{\Rb,\Cs}(\nu)  \nonumber \\
 &-g_{b}^{\Rb} g_{a}^{\Cs} S_{bx,ax}^{\Rb,\Cs}(\nu)  +g_{b}^{\Rb} g_{b}^{\Cs} S_{bx,bx}^{\Rb,\Cs}(\nu),\label{eq:model}
\end{align}
where the atom-light coupling factor $g_{\alpha}^{\beta}$ depends on the wavelength of the laser probing the $\alpha$ hyperfine spin of the $\beta$ species. We remind the reader that $S_{\alpha x,\alpha' x}^{\beta,\beta'}(\nu)$, given by Eq. \eqref{eq:CrossCorrelationSpectrum0}, corresponds to the cross-spectrum between the $\alpha$ and $\alpha'$ hyperfine spins of the $\beta$ and $\beta'$ species, respectively, measured along the $x$-axis.

Moreover, the model also requires as input the steady-state covariance matrix $R(0)$. We choose a diagonal $R(0)$, because as discussed in Appendix~\ref{sec:AppC}, if the imaginary part of the cross spectrum is zero and the spin-variances follow the scaling outlined in Eq. \eqref{eq:Ap:NoiseScaling}, then $R(0)$ must be diagonal. We have verified that the measured imaginary part of the cross spectrum is zero within the measurement resolution. We did so not only for the operating wavelengths of the two probe lasers, but for four different probe-wavelength pairs (combination of two different wavelengths for $^{133}$Cs and $^{87}$Rb probing). This way, and given the dependence of the coupling factors $g_{\alpha}^{\beta}$ on the probe light wavelength, we alter the contribution of each of the four terms of Eq. \eqref{eq:model}, showing that the zero imaginary part of the cross spectrum is not accidental, but reflects an underlying property of all cross-spectra appearing in Eq. \eqref{eq:model}. Additionally, assuming the aforementioned scaling of the spin variances, we arrive at the result that $R(0)$ is diagonal.

Both the single-species power spectra and the dual-species cross-spectrum are used for the optimization of the fitted parameters. That is, the parameters are adjusted to achieve a minimum in the merit function:
\begin{widetext}
\begin{equation}
\sum_{j=1}^{800} \Bigg \{ \left[ S_{\rm meas}^{\Rb,\Cs}(\nu_j)-S_{\rm model}^{\Rb,\Cs} (\nu_j;{\cal P}) \right]^2  + \left[ S_{\rm meas}^{\Rb,\Rb}(\nu_j)-S_{\rm model}^{\Rb,\Rb} (\nu_j;{\cal P}) \right]^2 +\left[ S_{\rm meas}^{\Cs,\Cs}(\nu_j)-S_{\rm model}^{\Cs,\Cs} (\nu_j;{\cal P} ) \right]^2 \Bigg \},
\end{equation}
\end{widetext}
where $S_{\rm meas}^{\beta,\beta '}(\nu_j)$ and $S_{\rm model}^{\beta,\beta '}(\nu_j;{\cal P})$ is the experimental and the model's prediction for the noise spectrum between spin species $\beta$ and $\beta '$, measured and calculated at the $j$-th frequency bin of the spectrum analyzer, respectively, where $j=1,2,...,800$. The symbol ${\cal P}$ represents the set of all fitted parameters, including the spin-exchange rates ($R^{\Rb,\Rb}_{\rm{se}}$, $R^{\Cs,\Cs}_{\rm{se}}$), the S-damping rate ($R$), the spin relaxation rate due to diffusion ($R_{\text{D}}$), the magnetic field ($B_{0}$), the photon shot noise levels (${\rm PSN}_{\Rb}$,  ${\rm PSN}_{\Cs}$), and the scaling factors ($\mathcal{K}_{\Rb}$, $\mathcal{K}_{\Cs}$) for each of the Rb and Cs power spectrum. The scaling factor of the Rb-Cs cross-spectrum is $\sqrt{\mathcal{K}_{\Rb} \mathcal{K}_{\Cs}}$ and does not appear as independent variable for the fitting. 

In Fig.~\ref{fig:CrossCorrelationData} we present the data (Rb-Cs cross-correlation spectrum and single-species power spectra) with the result of the global fit, showing very good agreement with the theoretical model. In the table shown in Fig. \ref{fig:CrossCorrelationData}m we summarize the fit parameters. The fitted value of $B_0$ follows the expected value based on the applied current to the coil within the magnetic shields, with a deviation of only a few percent at the smallest field values. The fitted values of the spin-exchange rates also agree within $10\%$ with the  values derived from the spin-exchange cross-sections reported in literature \cite{seltzerthesis}. Importantly, the fitted values for the spin-exchange rates, photon shot noise levels, and scaling factors were consistent across the six different magnetic field values, demonstrating the internal consistency of the theoretical model, which captures the global magnetic-field dependence of the data. Lastly, the S-damping and diffusion rates could not be accurately determined from the fit, likely because they are two orders of magnitude smaller than the spin-exchange rates. To address this issue, the fitting of those rates was constrained within a range spanning from a factor of 3 below to a factor of 3 above the expected rates based on the relevant experimental paarameters. 
\section{\label{sec:results} Results and Discussion}
 In the cross-correlation spectra (Figs. \ref{fig:CrossCorrelationData}a-f), a prominent positive peak is observed at low magnetic fields in the frequency range near the average of the resonance frequencies of the two species, as determined by their power spectra (Figs. \ref{fig:CrossCorrelationData}g-l). The peak height drops and becomes broader with increasing magnetic field. This behavior is primarily caused by the spin-exchange collisions between the alkali-metal atoms and reflects how the effect of these collisions changes with the magnetic field. 

The positive and negative swings of the cross-correlation spectrum indicate that band-limited measurements can exhibit either positive or negative correlations between the two spin species. In other words, the value and sign of the correlations depend on the bandwidth of the measurement and on the central measurement frequency. To illustrate this point, we consider the cosine quadratures $\widetilde{\mathcal{F}_\beta} = (1/T_{\text{BW}})\int_{0}^{T} \mathcal{F}_\beta (t)e^{(T-t)/T_{\text{BW}}}\cos (\Omega t) dt$ of the (collective, transverse) spin for each of species $\beta\in\{\Rb,\Cs\}$, and examine their correlation,  $\langle\widetilde{\mathcal{F}_\Rb}\widetilde{\mathcal{F}_\Cs}\rangle$, where $T$ is the measurement time, $\Omega$ is the frequency of the harmonic quadrature, and $1/(2 \pi T_{\text{BW}})$ is the integration bandwidth. We parenthetically note that such cosine (or sine) quadratures are typically measured in AC magnetometers with a lock-in amplifier. This correlation can be expressed as an integral of the inter-species cross-correlation spectrum, $\langle\widetilde{\mathcal{F}_\Rb}\widetilde{\mathcal{F}_\Cs}\rangle= \int_{0}^{\infty} S^{\Rb,\Cs}_{\rm meas}(\nu) \phi(\nu) d\nu$, where $\phi(\nu)$ is a kernel function that accounts for the effective measurement bandwidth and depends on the quadrature frequency $\Omega$ and the integration bandwidth $T_{\text{BW}}$, with a negligible effect from the measurement time $T$ when $T\gg T_{\text{BW}}$ and $\Omega T_{\text{BW}} \gg 1$ (see Appendix \ref{sec:app:BandwidthEffect} for an explicit formula for the kernel function).  We characterize the strength of this cross-correlation with the coefficient
\begin{figure}[th!]
\includegraphics[width=8cm]{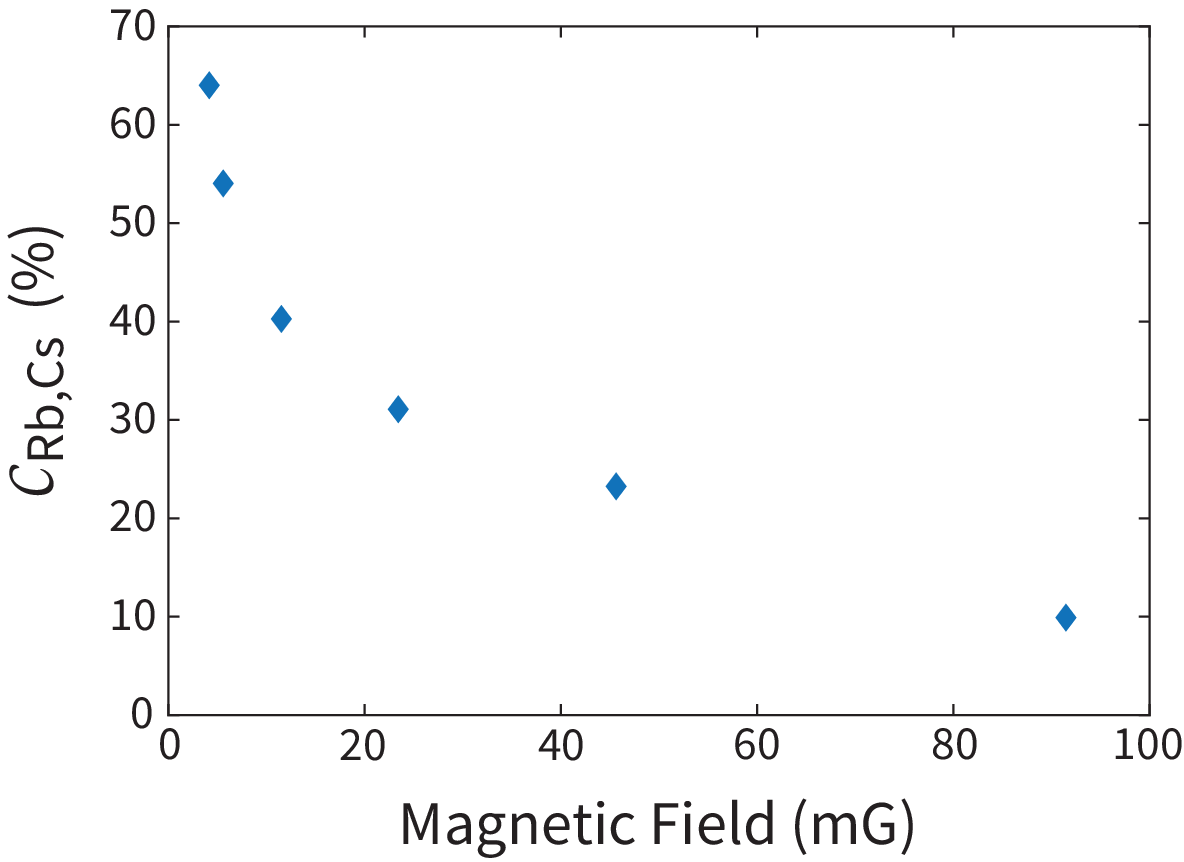}
\caption{Cross-correlation coefficient between \textsuperscript{87}Rb and \textsuperscript{133}Cs spins, both estimated at the frequency where the cross-correlation spectrum is maximum using a bandwidth of 100 Hz.}
\label{fig:CorrelationVsB}
\end{figure}
\small
\begin{equation}
{\cal C}_{\Rb,\Cs} = \frac{\int_{0}^{\infty} S_{\rm meas}^{\Rb,\Cs}(\nu) \phi_{\Rb\Cs}(\nu)d\nu}{\sqrt{\int_{0}^{\infty} S_{\rm meas}^{\Rb,\Rb}(\nu) \phi_\Rb(\nu) d\nu\int_{0}^{\infty} S_{\rm meas}^{\Cs,\Cs}(\nu) \phi_\Cs(\nu) d\nu}},\nonumber
\end{equation}
\normalsize
where the kernel functions $\phi_\Rb(\nu)$, $\phi_\Cs(\nu)$, and $\phi_{\Rb\Cs}(\nu)$ are centered around $\nu_0^{\Rb}$, $\nu_0^{\Cs}$ and $\nu_0^{\Rb\Cs}$, respectively, i.e. the frequency where the corresponding spectrum is maximum (additionally, all three kernel functions depend on the measurement time $T$). 

In Fig.~\ref{fig:CorrelationVsB} we plot ${\cal C}_{\Rb,\Cs}$ as a function of the magnetic field for a measurement bandwidth of 100~Hz, using the experimentally acquired spectra. It is seen that at low magnetic fields, the cross-correlation between the two spin-species can be a significant fraction of the measured spin-noise power, while it drops at larger magnetic fields. Alternatively, if the center frequencies of the harmonic quadratures are chosen in the region where the cross-spectrum is negative, the measured correlation (for appropriate integration time) will correspondingly appear to be negative. Overall, the sign and strength of the cross-correlation can be adjusted with readily controllable experimental parameters, like the magnetic field or the measurement bandwidth.
\subsection{Cross-correlation spin-noise power}
Of particular interest is the equal-time ($\tau=0$) cross-correlation, i.e. the total cross-correlation power. The power is related to the noise terms that enter into the stochastic differential equations \cite{gardiner2009stochastic} describing the time evolution of the observables. Previously, a debate about the value of cross-correlation power has emerged in the literature. In particular, Dellis and coworkers \cite{Dellis} used a \textsuperscript{85}Rb-\textsuperscript{87}Rb spin ensemble and measured a non-zero cross-correlation power, which increased at low magnetic fields. In constrast, Roy and coworkers \cite{roy2015cross} used a \textsuperscript{85}Rb-\textsuperscript{133}Cs spin ensemble and found the cross-correlation power to be zero, irrespective of the magnetic field. 
\begin{figure}[th]
\includegraphics[width=8cm]{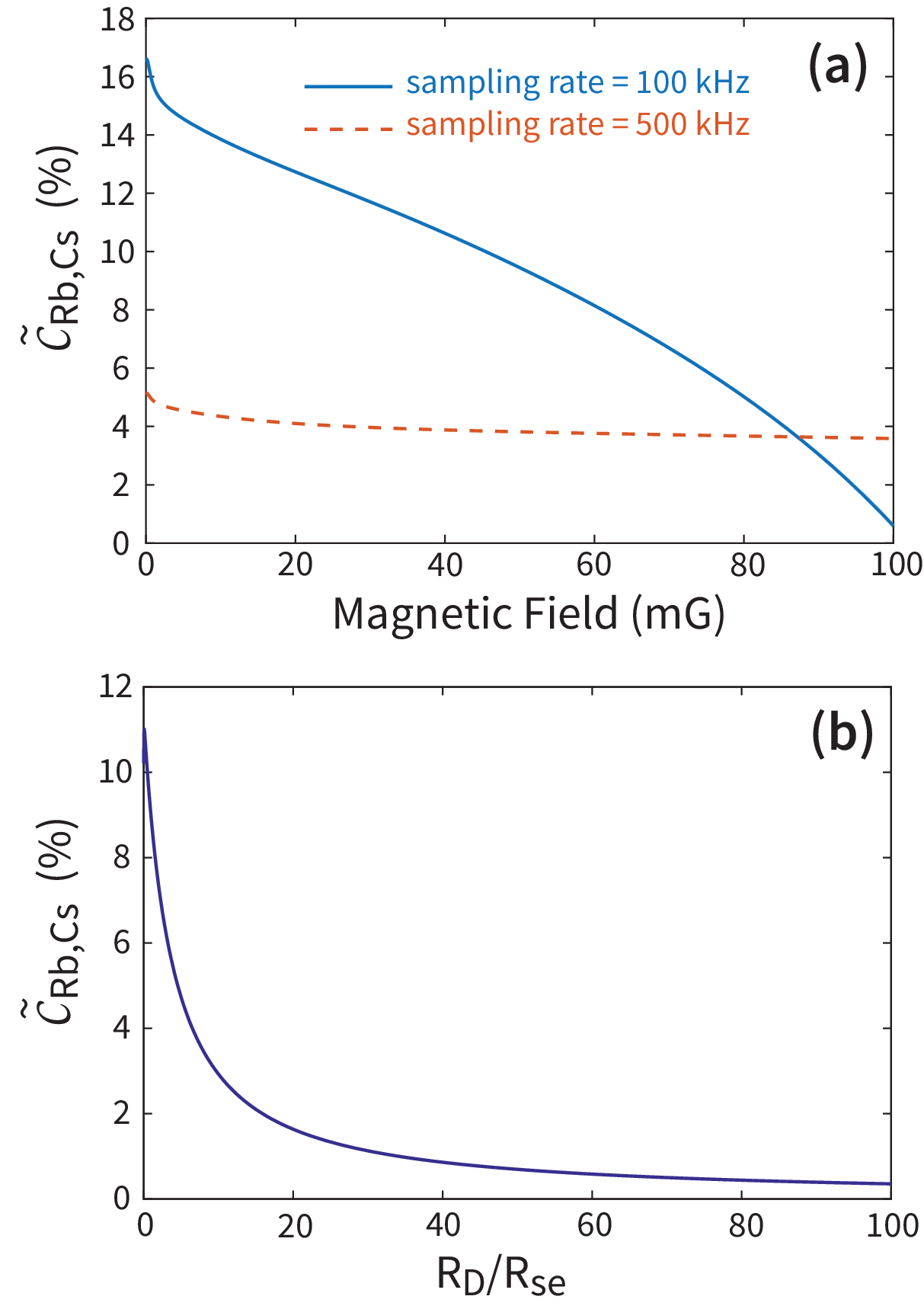}
\caption{(a) Cross-correlation coefficient as a function of the magnetic field for two different sampling rates. (b) Cross-correlation coefficient for a specific magnetic field, $B=10~{\rm mG}$, as a function of the ratio $R_D/R_{\rm se}$, quantifying the strength of transit-time-broadening relative to spin-exchange relaxation. The inter-species cross-correlation is suppressed if the linewidth is not dominated by spin-exchange relaxation.}
\label{noisepower}
\end{figure}

We will here resolve the aforementioned debate, noting that the measurement bandwidth has a subtle effect on the observed cross-correlation power. Firstly, as noted previously, a zero imaginary part of the cross-spectrum and a physically justifiable scaling of the single species spin-noise variance indeed imply a zero cross-correlation power. This, however, corresponds to the integration of the cross-correlation spectrum from frequency zero to infinity. In a realistic experiment, all measurements are conducted within a finite frequency range. Frequency components exceeding this range do not contribute to the measured cross-correlation power. Consequently, if the cross-correlation spectrum contains considerable power in the frequency range beyond the measurable bandwidth, the detected cross-correlation power may indeed appear to be non-zero. This is particularly the case when the high-frequency tail of the spectrum extends to frequencies larger than the spin-exchange rate.
To demonstrate this, we calculate the \textsuperscript{87}Rb-\textsuperscript{133}Cs cross-correlation coefficient when acquiring the spin signals at a finite sampling rate. Each recorded data point is modelled as the average of the corresponding signal over the sampling duration: $\tilde{y}_{\alpha} = {1\over {\Delta t}}\int_{t}^{t+\Delta t} y_{\alpha}(t') dt' $, where $\tilde{y}_{\alpha}$ and $y_{\alpha}$ represent respectively the sampled and the underlying continuous-time signals of species $\alpha$, and $1/\Delta t$ denotes the sampling rate. We define the cross-correlation coefficient $\tilde{\cal C}_{\Rb,\Cs}=\langle \tilde{y}_{\Rb} \tilde{y}_{\Cs} \rangle / \sqrt{\langle \tilde{y}^2_{\Rb}\rangle \langle \tilde{y}^2_{\Cs}\rangle}$, and calculate it from the theoretical fits to the data employing Eq.\eqref{eq:App:BWEffect41}.

The coefficient $\tilde{\cal C}_{\Rb,\Cs}$ is shown in Fig.~\ref{noisepower}a as a function of the magnetic field for two different sampling rates. It is seen that we obtain a significantly smaller correlation for a large sampling rate, as in this case a larger part of the negative high-frequency tail of the cross-correlation spectrum is included. These observations explain the positive correlations observed in \cite{Dellis}, where the bandwidth of the measurement was 50 kHz.

Regarding the experimental result in \cite{roy2015cross}, we note that a crucial physical behavior is the dependence of the correlation coefficient on the spin-exchange relaxation rates $R_{\rm se}$, and all other relaxation rates. In Fig. \ref{noisepower}b we plot the correlation coefficient as a function of the ratio $R_D/R_{\rm se}$, where $R_D$ is the relaxation due to atoms diffusing out of the probe beam (see Appendix 2.A.f). It is seen that for a large $R_D$ the correlation drops to zero. This is because for correlations to be observable, the magnetic linewidths should be dominated by spin-exchange broadening. If other broadening mechanisms prevail, the correlation effect will be suppressed. Diffusion out of the probe beam, or the so-called transit-time broadening, is indeed one such possibility, apparently responsible for the zero cross-correlation power measured in \cite{roy2015cross}. The authors in \cite{roy2015cross} focused the probe beam to a $50~{\rm \mu m}$ waist. Moreover, they had 3 times less buffer gas pressure than in this experiment. This renders the transit time broadening about 300 times larger than our case, amounting to $8.4\times 10^5~{\rm s}^{-1}$. A subtle difference of this kind of broadening is that it equally affects the nucleus and the electron, since it is just a Fourier broadening of the signal's limited-time observation. Hence, there is no slowing down factor like in the other broadening mechanisms, and the aforementioned rate directly appears in the measured linewidth. In fact, it can be seen from Fig. 2 of \cite{roy2015cross} that the linewidth is about 100 kHz, even though the cesium and rubidium densities reported in \cite{roy2015cross} are similar to our experiment, hence the spin-exchange broadening should be about 4 kHz. 
\subsection{Intra-species correlations in the context of mean field theory}
Using the measurements on two district species, it is clearly demonstrated that spin-exchange collisions result in unequal-time correlations among the atoms involved in collisions. Among those, there are also inter-atomic unequal-time correlations between atoms of the same species. It seems remarkable that a mean field theory employing essentially a single-atom description captures the intricate correlations emerging from spin-exchange collisions among atoms of identical species.

To address this issue and underscore the internal consistency of the presented theory, we proceed to demonstrate the reduction of the equations of motion governing two colliding atoms engaging in spin-exchange interactions within the same species. This reduction elegantly transforms the dynamics of the interacting pair into the equation of motion characterizing a mean single atom, which is effectively measured in an experiment. This elucidates the connections between the microscopic behavior of individual collisions and the macroscopic behavior described by mean field theory, thereby shedding light on the overarching consistency of the theoretical framework.

Consider the observable vector:
\begin{equation}
\mathcal{\hat{V}}(t)=\left(\hat{f}^\mathfrak{a}_{a,x},\hat{f}^\mathfrak{a}_{a,y},\hat{f}^\mathfrak{a}_{b,x},\hat{f}^\mathfrak{a}_{b,y},\hat{f}^\mathfrak{b}_{a,x},\hat{f}^\mathfrak{b}_{a,y},\hat{f}^\mathfrak{b}_{b,x},\hat{f}^\mathfrak{b}_{b,y} \right)^{\top},
\end{equation}
where now the indices $\mathfrak{a}$ and $\mathfrak{b}$ denote atoms of the same species. All spin operators depend on time $t$. Using the methods presented in Appendix \ref{sec:MeanSpinDynamics}, the mean spin dynamics can be formulated as follows:
\begin{equation}
\frac{d}{dt} \langle \mathcal{\hat{V}}(t)\rangle= \tilde{\mathcal{A}} \langle \mathcal{\hat{V}}(t)\rangle.
\end{equation}
where the matrix $\tilde{\mathcal{A}}$ encapsulates the linear dynamics akin to the representation by matrix $A$ pertinent to the dual-species case.

In an experiment, collective variables $\hat{f}^\mathfrak{a}+\hat{f}^\mathfrak{b}$ are measured. We thus define the collective observable vector
\begin{equation}
\begin{split}
{\cal \hat{M}}(t) & = \mathcal{R} \mathcal{\hat{V}}(t) \\
& =\left(\hat{f}^\mathfrak{a}_{a,x}+\hat{f}^\mathfrak{b}_{a,x},\hat{f}^{\mathfrak{a}}_{a,y}+\hat{f}^\mathfrak{b}_{a,y},\hat{f}^\mathfrak{a}_{b,x}+\hat{f}^\mathfrak{b}_{b,x},\hat{f}^\mathfrak{a}_{b,y}+ \hat{f}^\mathfrak{b}_{b,y}\right)^{\top},
\end{split}
\end{equation}
where $\mathcal{R}$ is the transformation matrix from the two-atom spin space to the reduced collective-spin space:
\begin{equation}
\mathcal{R} = \left(
\begin{array}{cccccccc}
 1 & 0 & 0 & 0 & 1 & 0 & 0 & 0 \\
 0 & 1 & 0 & 0 & 0 & 1 & 0 & 0 \\
 0 & 0 & 1 & 0 & 0 & 0 & 1 & 0 \\
 0 & 0 & 0 & 1 & 0 & 0 & 0 & 1 \\
\end{array}
\right).
\end{equation}

Simple matrix algebra verifies that the structure of matrix $\tilde{\mathcal{A}}$ is such that it fulfills the equality:
\begin{equation}
\mathcal{R} \tilde{\mathcal{A}} \mathcal{\hat{V}} = \mathcal{R} \tilde{\mathcal{A}} \mathcal{R}^{+} \mathcal{\hat{M}}, 
\end{equation} 
where $\mathcal{R}^{+}$ is the Moore-Penrose inverse \footnote{Since $\mathcal{R}$ has linearly independent rows, the Moore-Penrose inverse is the right inverse of $\mathcal{R}$ and: $\mathcal{R}^{+}=\mathcal{R}^{\dag} (\mathcal{R} \mathcal{R}^{\dag})^{-1}$. } of $\mathcal{R}$. As a result:
\begin{equation}
\frac{d}{dt}\langle {\cal \hat{M}}(t) \rangle = \mathcal{R} \frac{d}{dt} \langle \mathcal{\hat{V}} (t)\rangle =\mathcal{R} \tilde{\mathcal{A}} \mathcal{R}^{+} \langle \mathcal{\hat{M}}(t) \rangle =  \mathcal{A}\langle \mathcal{\hat{M}}(t) \rangle,
\end{equation}
where $\mathcal{A}=\mathcal{R} \tilde{\mathcal{A}} \mathcal{R}^{+}$. 

The matrix $\mathcal{A}$, formulated to describe the evolution of the collective ensemble spin, corresponds precisely to the matrix derived through the utilization of the equation governing the density matrix of a single atom (see for example Eq.~45 of \cite{appelt}). This congruence implies that the single-atom equation implicitly encompasses the influence of dynamic (non-equal time) inter-atomic correlations emanating from collisions involving atoms of the identical species.
\subsection{\label{sec:ent}Character of correlations: entanglement}
As mentioned in the introductory Sec. IIB, a pertinent question is whether the observed multi-time correlations are quantum or classical in nature. In other words, whether the observed correlations can be described using a classical probability model or are they non-classical in the sense that they cannot be prepared with classical operations. Quantifying quantum or quantum/classical correlations \cite{adesso2016measures} can be rather challenging and goes beyond the scope of this work. We briefly note that quantum correlations comprise an entire family of relationships \cite{adesso2016measures}, such as entanglement and quantum discord, which could potentially serve as benchmarks for the observed correlations. 

We here make only some exploratory comments and leave a more detailed discussion for future work. In \cite{PhysRevA.103.L010401} it was shown that spin-exchange collisions of the kind encountered in our work can create entanglement between two (partially) polarized spin ensembles that can be sustained for meaningful timescales. Similarly, it was shown in \cite{Ofer2, Ofer1} that spin-exchange collisions under appropriate conditions create quantum correlations and can be harnessed to transfer the quantum state from one spin species to another spin species.  

On the other hand, as explained in the previous subsection, the correlations measured here are consistent with the theory describing spin dynamics from the single-atom perspective (mean-field), hence such correlations cannot be anything but classical. It is possible, however, that the character of correlations could have a parametric dependence on the spin-polarization of the vapor, like the Werner state discussed in the introduction. We here use a simple two-atom toy model to illustrate that this seems might indeed be the case. Considering atoms spin-polarized along $\mathbf{\hat{x}}$, a magnetic field along $\mathbf{\hat{z}}$ creates a spin component along $\mathbf{\hat{y}}$. What we will show is that we expect the fluctuations of the total spin component $\hat{f}_y$ for \textsuperscript{133}Cs and \textsuperscript{87}Rb to be quantum correlated. 

\begin{figure}[th!]
\includegraphics[width=8cm]{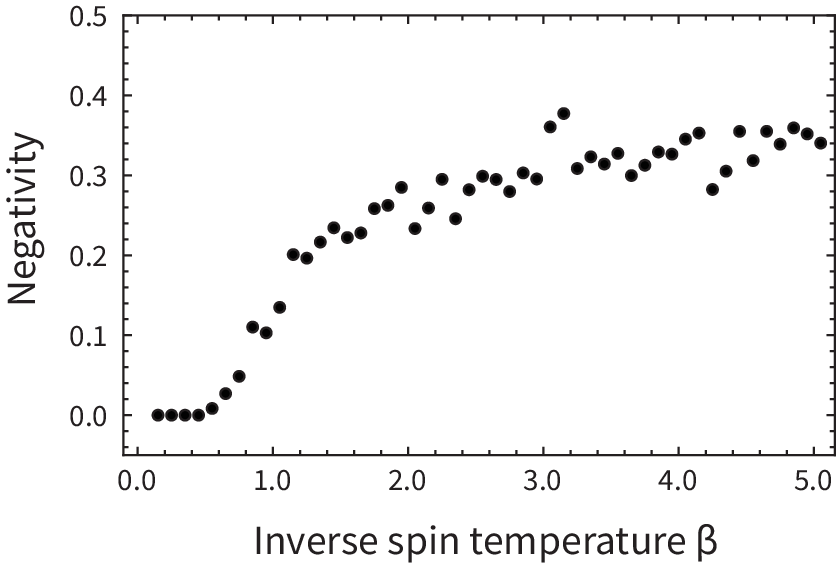}
\caption{Negativity quantifying the entanglement of the \textsuperscript{87}Rb-\textsuperscript{133}Cs combined spin state produced by (i) spin-polarizing the vapors along the $x$-axis with equal and opposite polarizations, (ii) spin precession in a transverse magnetic field, and (iii) cross-spin exchange collisions.} 
\label{fig:ent}
\end{figure}

We take as initial states for the \textsuperscript{133}Cs and \textsuperscript{87}Rb atoms the spin-temperature states $\rho_1=e^{\beta \hat{f}_x}/\tr\{\beta \hat{f}_x\}$ and $\rho_2=e^{-\beta \hat{f}_x}/\tr\{-\beta \hat{f}_x\}$, i.e. having the same but opposite spin polarization $\braket{\hat{s}_x}={1/ 2}\tanh{({\beta/ 2})}$, with $\beta$ being the spin temperature \cite{appelt}. The combined initial state is $\rho=\rho_1\otimes\rho_2$. Note that for each atom the operator $\hat{f}_x$ and the respective density matrix has a matrix representation of different dimension. We then evolve $\rho$ by the Hamiltonian $\hat{H}$, i.e. we calculate $\rho'=e^{-i\hat{H}t}\rho e^{i\hat{H}t}$, where $\hat{H}=\hat{h}_{\rm Cs}\otimes\mathbb{1}+\mathbb{1}\otimes \hat{h}_{\rm Rb}$, with $\hat{h}_{\rm Cs}$ and $\hat{h}_{\rm Rb}$ being the \textsuperscript{133}Cs and \textsuperscript{87}Rb Breit-Rabi Hamiltonians, respectively. We use a small field of $10~{\rm \mu G}$ and a precession time of $10~{\rm \mu s}$. We then apply on $\rho'$ the spin-exchange operator $\hat{P}_e=(1/2)\mathbb{1}+2 \mathbf{\hat{s}}_1 \cdot \mathbf{\hat{s}}_2 $, and calculate the resulting negativity \footnote{To quantify the entanglement in the multi-dimensional Hilbert space of the two alkali species we use the negativity criterion \cite{10.5555/2011706.2011707} (entanglement measure) as demonstrated in \cite{PhysRevA.103.L010401}.} of the state $\hat{P}_e \rho' \hat{P}_e^{\dagger}$, like in \cite{PhysRevA.103.L010401}. The result is shown in Fig. \ref{fig:ent}. 
It is seen that for low polarizations pertinent to the noise measurement of this work the two atoms are not entangled. However, there appears to be a threshold spin-polarization over which the two atoms gradually become strongly entangled. 
\section{\label{sec:conc} Conclusions}
We have studied experimentally and theoretically spin correlations that spontaneously build up in a dual-species alkali-metal vapor, here composed of $^{87}{\rm Rb}$ and $^{133}{\rm Cs}$. We have identified and categorized various types of spin-correlations that can be probed in a hot atomic vapor and we have elucidated their behavior. The combined action of interspecies and intraspecies spin-exchange collisions leads to positive equal-time spin-correlations which are enhanced at low magnetic fields and suppressed at high fields due to an interplay of their unique spectral distribution and the measurement bandwidth. 

The nature of these correlations has been discussed, anticipating that similar correlations in spin-polarized vapors are expected to be genuinely quantum, i.e. to reflect interspecies spin entanglement. 

The use of two species rather than one, and the study of both auto- and cross-correlations, helps to unravel the complexity of spin dynamics in alkali-metal vapors, so far treated mostly as consisting of uncorrelated atoms, and can have significant repercussions for the field of quantum metrology with hot atomic vapors. 
\acknowledgements
K.M. acknowledges support from Grant FJC2021047840-I funded by MCIN/AEI/10.13039/501100011033; the European Union ``NextGenerationEU/PRTR''; Greece and the European Union [European Social Fund (ESF)] through the Operational Programme ``Human Resources Development, Education and Lifelong Learning'' in the context of the project ``Strengthening Human Resources Research Potential via Doctorate Research'' (Grant No. MIS-5000432), implemented by the State Scholarships Foundation (IKY). GV acknowledges funding from EU QuantERA Project PACE-IN (GSRT Grant No. T11EPA4-00015) and from the Hellenic Foundation for Research and Innovation (HFRI) under the HFRI agreement  no. HFRI-00768 (project QCAT). F.V., A.M., G.M., M.S., G.P.T. and I.K.K. acknowledge the cofinancing of this research by the European Union and Greek national funds through the Operational Program Crete 2020-2024, under the call ``Partnerships of Companies with Institutions for Research and Transfer of Knowledge in the Thematic Priorities of RIS3Crete", with project title ``Analyzing urban dynamics through monitoring the city magnetic environment'' (project KPHP1 - 0029067). M.W.M acknowledges the European Commission project OPMMEG (101099379), Spanish Ministry of Science MCIN with funding from European Union NextGenerationEU (PRTR-C17.I1) and by Generalitat de Catalunya ``Severo Ochoa'' Center of Excellence CEX2019-000910-S; projects SAPONARIA (PID2021-123813NB-I00) and MARICHAS (PID2021-126059OA-I00) funded by MCIN/ AEI /10.13039/501100011033/ FEDER, EU; Generalitat de Catalunya through the CERCA program;  Ag\`{e}ncia de Gesti\'{o} d'Ajuts Universitaris i de Recerca Grant 2021-SGR-01453; Fundaci\'{o} Privada Cellex; Fundaci\'{o} Mir-Puig;. A.K. and M.L. were supported by the Hellenic Foundation for Research and Innovation (H.F.R.I.) under the “First Call for H.F.R.I. Research Projects to support Faculty members and Researchers and the procurement of high-cost research equipment grant” (Project ID HFRI-FM17-1034). M.S. acknowledges financial support from the Spanish Agencia Estatal de Investigac\'{i}on, Grant No. RYC2021-032032-I, PID2019107609GB-I00, and the Catalan Government for the project QuantumCAT 001-P-001644, co-financed by the European Regional Development Fund (FEDER).
\appendix
\section{\label{sec:AppA}Detailed Theoretical Derivation of Cross- and Auto-Correlation Spectra}
We here recapitulate the basic physics of our theoretical description, namely atom-light coupling, and atomic evolution due to coherent Hamiltonian dynamics and relaxation effects dominated by atom-atom collisions. 
\subsection{Atom-light coupling}
The interaction of light with atomic spin ensembles has been described in detail elsewhere \cite{PhysRev.163.12,stockton,Deutsch2010,PhysRevA.77.032316}. For near-resonant monochromatic light of non-saturating intensity, the coherent atom-light interaction (i.e. the interaction that leads to forward scattered light in the spatial mode of the probe beam) is described with a polarizability Hamiltonian that couples the magnetic sublevels of the atomic ground state to the polarization modes of the light. The resulting atomic polarizability is a rank-2 tensor operator that can be decomposed into three irreducible components. Of those, the tensor (rank-2) polarizability is negligible for the conditions of our experiment (pressure broadening $\sim 10$~\si{\giga\hertz} and detuning $\sim 100$~\si{\giga\hertz}). Most relevant for our experiment is the vector polarizability (rank-1) describing paramagnetic Faraday rotation and reading \cite{comParaxModes}:
\begin{equation}
\hat{H}_{\text{int}} = \sum_{i=1}^{\NAtoms} \hat{{\cal S}}_3(\mathbf{r}_i,t) \beta(\mathbf{r}_i) \left[ g_a \hat{f}_{a,x}^{(i)}(t) - g_b \hat{f}_{b,x}^{(i)} (t)\right],
\label{eq:Hint}
\end{equation}
where the summation is performed over all atoms probed by the laser beam, ${\cal S}_3(\mathbf{r},t)=i \left( \hat{a}^{\dag}_{y}\hat{a}_{z}-\hat{a}^{\dag}_{z}\hat{a}_{y} \right)/2$ is the Stokes light-operator quantifying the photon-flux imbalance of the left- and right- circular polarization modes (and in the most general case is a function of space and time), $\mathbf{r}_i$ is the location of the $i$-th atom, $\hat{f}_{a,x}^{(i)}(t)$ and $\hat{f}_{b,x}^{(i)}(t)$ are dimensionless spin components of the $i$-th atom along the probe laser direction in the $a=I+1/2$ and $b=I-1/2$  hyperfine levels of the ground state, with $I$ being the nuclear spin quantum number ($I=7/2$ for $^{133}$Cs and $I=3/2$ for $^{87}\Rb$). The parameter $\beta(\mathbf{r}_i)$  characterizes the local field intensity associated with the spatial mode of the probe beam (see below).

For a D2 optical transition linewidth dominated by pressure broadening the coupling constants entering Eq.\eqref{eq:Hint} are approximated by \cite{MitchellNatureCom}:
\begin{equation}
g_{\alpha} \approx \frac{1}{2I+1} \frac{c r_{\text{e}} f_{\text{osc}}}{A_{\text{eff}}} \frac{\nu_{\ell}-\nu_{\alpha}}{(\nu_{\ell}-\nu_{\alpha})^2+(\Delta \nu/2)^2},
\end{equation}
where $r_{\text{e}}\approx 2.82 \times 10^{-15}$~\si{\meter} is the classical electron radius, $f_{\text{osc}}$ is the oscillator strength of the optical transition, $c$ is the speed of light, $\Delta \nu$ is the pressure-broadened optical linewidth (FWHM), $\nu_{\ell}$ is the frequency of light, and $\nu_{\alpha}$ with $\alpha \in\{a,b\}$ is the resonance frequency of the corresponding ground-state hyperfine level. If $A_{\text{eff}}$ labels the effective area of the beam, it is $\beta(\mathbf{r}_i)/A_{\text{eff}} =I(\mathbf{r}_i)/\int I(\mathbf{r}_i) dy dz $, where $I(\mathbf{r}_i)$ is the light intensity at the coordinate $\mathbf{r}_i$ and $\int I(\mathbf{r}_i) dy dz$ is the total power of the beam. For a $\text{TEM}_{00}$ Gaussian beam it is $\beta(\mathbf{r}_i)/A_{\text{eff}} = e^{-2|\mathbf{r}_{i}|^2/w(x)^2}/(\pi w(x)^2/2)$, where $w(x)$ is the $x$-dependent Gaussian beam width.

For a  linearly polarized probe in the $y$-direction, neglecting the time it takes light to propagate through the spin ensemble \cite{PhysRevA.71.032348}, the solution of the Heisenberg equation of motion for the light operator (in the case of small rotation angles) yields \cite{Kuzmich}:
\begin{equation}
\hat{{\cal S}}_2^{(\text{out})} (t) \approx \hat{{\cal S}}_2^{(\text{in})} (t)+ \frac{1}{2}\Phi \bar{g} \mathcal{\hat{F}} (t)\label{eq:S2out_Faraday},
\end{equation} 
where $\Phi$ is the photon-flux (photons per time). The out (in) superscripts denote the operator after (before) interacting with the atomic sample, $\bar{g}= (|g_a|+|g_b|)/2$ is the mean coupling constant to the two ground-state hyperfine manifolds and $\mathcal{\hat{F}} (t)$ is the measured collective spin defined as:
\begin{align}
\mathcal{\hat{F}}(t) & \equiv \sum_{i=1}^{\NAtoms}\beta(\mathbf{r}_i) \left[ \frac{g_a}{\bar{g}} \hat{f}_{a,x}^{(i)} (t) - \frac{g_b}{\bar{g}} \hat{f}_{b,x}^{(i)} (t) \right]
\label{eq:collective_spin} \\
& = \int d \mathbf{r}  \beta(\mathbf{r})   \left[ \frac{g_a}{\bar{g}} \hat{\tilde{f}}_{a,x} (\mathbf{r},t) - \frac{g_b}{\bar{g}} \hat{\tilde{f}}_{b,x} (\mathbf{r},t) \right]. \label{eq:collective_spin_density}
\end{align}
In the last equation the spin-density operator $\hat{\tilde{\mathbf{f}}}$ was introduced, which for the $\alpha$ hyperfine manifold is defined as:
\begin{equation}
\hat{\tilde{\mathbf{f}}}_{\alpha}(\mathbf{r},t) = \sum_{i=1}^{\NAtoms} \hat{\mathbf{f}}^{(i)}_\alpha(t) \delta(\mathbf{r}-\mathbf{r}_i).
\end{equation}



\subsection{Atomic evolution}
As will become apparent in the following, the measured fluctuation spectrum depends on the dynamics of the collective atomic spin in the ground state, which are determined by the atomic Hamiltonian, atomic collisions, atomic thermal motion, and the interaction with light. We next treat such atomic dynamics.
\subsubsection{Hamiltonian}
The electronic-ground-state Hamiltonian for an alkali-metal atom in the presence of a magnetic field $\mathbf{B}$ is:
\begin{equation}
\hat{H}_g = A_{\text{hf}} \hat{\mathbf{I}} \cdot \hat{\mathbf{s}}+g_s \mu_B \hat{\mathbf{s}} \cdot \mathbf{B},
\end{equation}
where $A_{\text{hf}}$ is the isotropic hyperfine coupling coefficient, $\hat{\mathbf{I}}$ and $\hat{\mathbf{s}}$ are the nuclear and electron spin operators, respectively,   $g_s\approx 2$ is the electron's g-factor, and $\mu_B \approx 9.27\times 10^{-24}$~J/T is the Bohr magneton. 
\subsubsection{Spin-exchange collisions}
Spin-exchange collisions are dominant in the physics of atomic spin state evolution. The spin-exchange interaction originates from the difference between the lowest singlet and triplet potential energy of the molecular system formed during a pairwise collision. The established formalism captures the effect of spin-exchange collisions with a mean-field density matrix, which for the $i$-th atomic species reads \cite{Happer_Book}:
\begin{equation}
\begin{split}
d\rho_\beta/dt  &= \sum_{\gamma} R_{\rm se}^{\beta,\gamma} \big( \phi_\beta \left (1+4 \langle \hat{\mathbf{s}}_\gamma \rangle \cdot \hat{\mathbf{s}}_\beta \big ) - \rho_\beta \right )\\
&-\sum_{\gamma}2 i R_{\rm se}^{\beta,\gamma}\kappa_{\beta\gamma} \left[ \langle \mathbf{s}_\gamma \rangle \cdot \mathbf{\hat{s}}_\beta , \rho_\beta  \right],  \label{eq:SpinExchangeDensityMatrixEq}
\end{split}
\end{equation}
where $\beta,\gamma \in\{\Rb,\Cs\}$, $\phi =\rho/4+\hat{\mathbf{s}} \cdot \rho \hat{\mathbf{s}}$ is the part of density matrix without electron polarization, $\langle \mathbf{\hat{s}}_\gamma \rangle = \text{Tr}\left[ \mathbf{\hat{s}} \rho_\gamma \right]$, $ \kappa_{\beta\gamma}$ is a dimensionless parameter characteristic for the mean-field interaction between the $\beta$ and $\gamma$ species, and $R_{\rm se}^{\beta,\gamma}\approx  n_\gamma \upsilon_{\rm th}^{\beta,\gamma} \sigma_{\rm se}^{\beta,\gamma}$ is the spin-exchange rate, given in terms of the atomic density $n_\gamma$ of species $\gamma$, the relative thermal velocity between the colliding partners $\upsilon_{\rm th}^{\beta,\gamma}$ and the spin-exchange cross-section $\sigma_{\rm se}^{\beta,\gamma}$. In Eq.\eqref{eq:SpinExchangeDensityMatrixEq} the summation runs over all the different species present in the vapor, including the species $\beta$. As we are dealing with unpolarized vapors, the effect of the frequency shift proportional to the parameter $ \kappa_{\beta\gamma}$ will be ignored in the following.

We note that Eq. \eqref{eq:SpinExchangeDensityMatrixEq} is a mean field theory. The spin-exchange interactions that an atom can experience are replaced with an effective average interaction and the spin-degrees of freedom of the different atoms are traced out to get a reduced mean density matrix representing the evolution of the ensemble. Essentially, Eq. \eqref{eq:SpinExchangeDensityMatrixEq} provides for coarse-grained dynamics over a length scale much larger than the mean-free path. Even in this case, Eq. \eqref{eq:SpinExchangeDensityMatrixEq} can capture correlations that can be generated by the dynamics, at least at the level of classical correlations (see Sec. \ref{sec:ent}).
\subsubsection{S-damping collisions}
Binary collisions between alkali atoms or between alkali-metal atoms and buffer gas (without spin) lead to S-damping \cite{appelt}:
\begin{equation}
d\rho/dt = R\left( \phi-\rho \right), \label{RelSD}
\end{equation} 
where the part of the density matrix with electron polarization is destroyed while the purely nuclear polarization remains unaffected. The S-damping rate, $R$, is orders of magnitude smaller than the spin-exchange rate, $R_{\rm{se}}$.
\subsubsection{Relaxation due to optical fields}
In \cite{appelt}, it is shown that for the experimentally relevant case of fast quenching (when excited atoms are much more likely to be quenched than to radiate a photon) and excited state $J$-damping rapid with respect to the hyperfine interaction, the net evolution of the single-atom density matrix due to optical (depopulation and repopulation) pumping by linearly polarized light is of the form \eqref{RelSD}, with $R$ given by  $\Phi \sigma_{\text{op}}$, the mean pumping rate per (unpolarized) alkali-metal atom given in terms of the absorption cross-section $\sigma_{\text{op}}$ and the photon flux $\Phi$. In principle, the effect of light on the atomic-spin evolution can be made negligible, either by using low photon flux (low power or large beam area) or by choosing large detuning. However, as seen from Eq.~\eqref{eq:AutoCorrelationPolarimetry}, this results in a reduction of the measured spin-noise to photon shot noise ratio, leading to an increase in the uncertainty of estimation of the spin-noise spectrum \footnote{In a noisy process, the uncertainty (standard deviation) in the periodogram estimate at a frequency is equal to the its expectation value at this frequency \cite{wirsching2006random}.} for a given number of repetitions. In practice, the laser power is chosen so as to optimize the measurement's signal-to-noise ratio. For our case in particular, the laser power should be such that spin relaxation by the optical fields is not dominant, i.e. spin dynamics should be dominated by spin-exchange collisions.
\subsubsection{Combined description}
Overall, the density matrix evolution for species $\beta$ reads
\begin{align}
d\rho_{\beta}/dt &= A_{\text{hf},\beta} \hat{\mathbf{I}}_{\beta} \cdot \hat{\mathbf{s}}_{\beta}+g_s \mu_B \hat{\mathbf{s}}_{\beta} \cdot \mathbf{B}+R\left( \phi_{\beta}-\rho_{\beta} \right) \nonumber\\ 
& +\sum_{\gamma} R_{\rm se}^{\beta,\gamma} \left\{ \phi_{\beta} \left (1+4 \langle \hat{\mathbf{s}}_{\gamma} \rangle \cdot \hat{\mathbf{s}}_{\beta} \right ) - \rho_{\beta} \right \},  
\label{eq:TotalDensityMatrixEvolution}
\end{align} 
where the relaxation rate $R$ includes all the relaxation processes, other than the spin-exchange collisions, which destroy electron polarization without affecting the nucleus.
\subsubsection{Atomic diffusion}
In a cell with buffer gas, as in our experiment, the atomic thermal motion becomes diffusive via velocity-changing collisions. The effect of diffusion on spin-noise measurements has been studied in various works. In \cite{Diffusion} the correlation function of diffusion-induced noise in unconfined systems is derived. A treatment based on the Bloch-Heisenberg-Langevin formalism is developed in \cite{Ofer2020}, also considering boundary conditions on the cell walls. In our experiment the probe beam diameter is much smaller than the cell dimensions, thus the results of \cite{Diffusion} are more relevant. There, it is shown that for diffusion through a $\text{TEM}_{00}$ Gaussian and well-collimated beam (Rayleigh range much larger than the cell length) the spin-noise spectrum of an atomic ensemble with transverse spin-relaxation rate $1/T_2$ and spin precession frequency $\nu_{\text{L}}$ is given by 
\begin{equation}
S(\nu) \propto \frac{w^2}{2 D} \Re \left[ e^{s} \Gamma(s) \right], \text{ } s = \frac{w^2}{4 D} \left[1/T_2 + 2 \pi i (\nu - \nu_{\text{L}})\right],
\end{equation}
where $D$ is the atomic diffusion constant, $w$ is the waist of the beam and $\Gamma(s) = \int_{s}^{\infty} (e^{-x}/x) \, dx$ is the incomplete gamma function. When $w^2/4 D T_2 \gg 1$ it is $e^{s} \Gamma(s) \approx (1-1/s)/s$, and the spectrum takes the form:
\begin{equation}
S(\nu) \sim 2\frac{1/T_2'}{(1/T_2')^2+4 \pi^2 (\nu - \nu_{\text{L}})^2} +\mathcal{O}\left( \frac{4 D T_2}{w^2} \right),
\end{equation}
where $1/T_2'=1/T_2+R_{D}$, with $R_{D} \sim 4 D/w^2$.
Therefore, when the diffusion time across the beam is much longer than the spin coherence time, the effect of diffusion on the spin-noise spectrum can be captured to a good approximation by introducing an additional relaxation term with rate $R_{D}$.

\subsection{Correlations}
From Eqs.~\eqref{eq:cross-corr} and \eqref{eq:S2out_Faraday} it follows that the cross-correlation measured in the experiment is:
\begin{align}
C_{\Rb,\Cs}(\tau) & = \frac{q_e^2 \eta_{\Rb} \eta_{\Cs} \Phi_{\Rb} \Phi_{\Cs} \bar{g}_{\Rb} \bar{g}_{\Cs}}{2} \nonumber \\
 &\times \langle \mathcal{\hat{F}}_{\Rb}(t+\tau) \mathcal{\hat{F}}_{\Cs}(t)+ \mathcal{\hat{F}}_{\Cs}(t) \mathcal{\hat{F}}_{\Rb}(t+\tau) \rangle,
 \label{eq:cross-species corr}
\end{align}
where we have taken into account that the polarization properties of the two beams before the interaction with the atoms are uncorrelated e.g. $\langle {\cal S}^{(\rm{in})}_{2,\Rb}(t) {\cal S}^{(\rm{in})}_{2,\Cs}(t')\rangle =0$. The single-channel auto-correlation function is:
\begin{equation}
C_{\beta,\beta}(\tau) =  2 q_e^2 \eta_{\beta} \Phi_\beta \Big[ \delta(\tau)+ \frac{1}{2} \eta_{\beta} \Phi_\beta \bar{g}^2_{\beta} \langle \mathcal{\hat{F}}_{\beta}(t+\tau)  \mathcal{\hat{F}}_{\beta}(t) \rangle  \Big], \label{eq:AutoCorrelationPolarimetry}
\end{equation}
where $\beta \in\{\Rb,\Cs\}$. In deriving Eq. \eqref{eq:AutoCorrelationPolarimetry} we assumed perfectly coherent light before the interaction with the atoms, i.e. $\langle {\cal S}^{(\rm{in})}_{2,\beta}(t) {\cal S}^{(\rm{in})}_{2,\beta}(t')\rangle = (\Phi_\beta /2) \delta(t-t')$. The detector inefficiency was modelled by a perfect detector preceded by a beam splitter of power transmissivity $\eta_\beta$ \cite{gardiner2004quantum}. The first term in Eq.\eqref{eq:AutoCorrelationPolarimetry} describes photon shot-noise, while the second contains information about the collective spin correlations. We remark that although the photon shot-noise does not explicitly appear in the inter-species correlator of Eq.\eqref{eq:cross-species corr}, it does contribute to the uncertainty of its estimation. In Eqs.\eqref{eq:cross-species corr} and \eqref{eq:AutoCorrelationPolarimetry} it was assumed that there is no correlation between the light and spin operators, implying that the spin variables remain unaffected by the polarization fluctuations of the input field. This is true for our experiment probing unpolarized spin ensembles with light (refer to section \ref{sec:LightShift}). However, this assumption breaks down when dealing with non-zero spin polarization.

As long as the optical rotation remains small, Eqs. \eqref{eq:S2out_Faraday}-\eqref{eq:AutoCorrelationPolarimetry} remain valid even for optically thick ensembles with significant light absorption. In such cases, the variable $\Phi$ represents the photon flux at the ensemble output, which can be significantly reduced compared to the photon flux at the ensemble input \cite{note:ThickEnsemble}.

Combing Eqs. \ref{eq:collective_spin_density} and \ref{eq:cross-species corr}, the measured correlation can be written as:
\begin{widetext}
\begin{equation}
C_{\Rb,\Cs}(\tau) = \mathcal{K} \int \int d\mathbf{r} d\mathbf{r}' \beta^{\Rb} (\mathbf{r}) \beta^{\Cs} (\mathbf{r}') \times \sum_{\alpha, \alpha'} g_{\alpha}^{\Rb}  g_{\alpha'}^{\Cs}  \left[ \langle \hat{\tilde{f}}_{\alpha,x}^{\Rb}(\mathbf{r},t+\tau)  \hat{\tilde{f}}_{\alpha',x}^{\Cs}(\mathbf{r}',t) \rangle +\langle \hat{\tilde{f}}_{\alpha',x}^{\Cs}(\mathbf{r}',t) \hat{\tilde{f}}_{\alpha,x}^{\Rb}(\mathbf{r},t+\tau)   \rangle \right],  \label{eq:crossCorrelationSpinDensity}
\end{equation}
\end{widetext}
where $\mathcal{K}$ is an overall scaling factor. A similar equation can be written for the autocorrelation function.
\subsection{Collective spin operators}
The dynamical evolution of spin correlations can be found from the quantum regression theorem (QRT) \cite{gardiner2004quantum}, which states that the spin-correlation functions follow the same equations of motion as the mean spins. Consequently, to explore the dynamic evolution of the collective spin correlations, it suffices to determine the mean dynamics of the corresponding spins. In this regard, the mean field equation (Eq.~\eqref{eq:TotalDensityMatrixEvolution}) can be utilized. We note that the validity of Eq.~\eqref{eq:TotalDensityMatrixEvolution} to describe the dynamics of the collective spin has been confirmed in numerous experiments. Remarkably, the QRT allows us to move deeper into the understanding of spin dynamics and address inter-atomic correlations. This is because the QRT connects the mean dynamics of the collective spin to the collective correlations, and the latter include a multitude of inter-atomic correlation terms. 


In the following, we derive a simplified version of Eq.~\ref{eq:crossCorrelationSpinDensity} for the measured cross-correlation.
We adopt the approach presented in \cite{Ofer1}, involving the definition of coarse-grained continuous local-symmetric spin-density operators. These coarse-grained operators are obtained through the spatial convolution of the spin-density operators with a window function, which remains non-zero over a specific volume (e.g. a Heaviside function). This volume must be adequately large to encompass a significant number of atoms while being sufficiently small to ensure a uniform interaction of all atoms with the probe beam.

As discussed in \cite{Ofer1}, the inherent stochastic nature of collision parameters and random pairings of colliding atoms introduces noise. However, remarkably, these noise terms do not significantly influence the temporal unfolding of correlations, affecting only the zero-time correlation. The coarse-grained spin-density operators evolve as a single entity within the chosen coarse-grained volume, and their mean values follow the mean-field equation over time.

In our experimental setup, characterized by a buffer gas pressure of 330 Torr and a large beam diameter, the coarse-grain can be selected to be adequately large; this ensures that atomic motion does not result in significant correlations between different coarse-grains. Correlations resulting from collisions are confined solely within the same coarse-grained region. As explained in \cite{Ofer1}, this leads to correlations in the coarse-grained spin-density operators that are proportional to the coarse-grained delta function.

Overall, disregarding the minor variation in the effect of spin-relaxation caused by the probe beam along the atomic ensemble, the measured correlation in the experiment can be expressed as (see Eq.~\ref{eq:crossCorrelationSpinDensity}):
\begin{align}
C_{\Rb,\Cs}(\tau) \propto \sum_{\alpha, \alpha'} g_{\alpha}^{\Rb}  g_{\alpha'}^{\Cs}  \Big[ & \langle \hat{\breve{f}}_{\alpha,x}^{\Rb}(t+\tau)  \hat{\breve{f}}_{\alpha',x}^{\Cs}(t) \rangle + \nonumber \\
&\langle \hat{\breve{f}}_{\alpha',x}^{\Cs}(t)  \hat{\breve{f}}_{\alpha,x}^{\Rb}(t+\tau)   \rangle \Big]. \label{eq:crossCorrelationSpinDensity2}
\end{align}
In the above equation $\hat{\breve{f}}^{\beta}_{\alpha,x}$ denotes the coarse-grained, local-symmetric spin-density operator and is related to the single-atom spin $\hat{f}$ by
\begin{equation}
\langle \hat{\breve{f}}^{\beta}_{\alpha,x} \rangle = n_{\beta} \langle \hat{f}^{\beta}_{\alpha,x} \rangle  \label{eq:densityOperatorSingleAtomSpin}
\end{equation} 
where $n_{\beta}$ is the atomic density of the species $\beta$. The equations of motion for the spin-density operators can be accordingly found from Eqs.~\eqref{eq:densityOperatorSingleAtomSpin} and \eqref{eq:TotalDensityMatrixEvolution}.
\section{\label{sec:AppMatrixA}Mean spin dynamics} \label{sec:MeanSpinDynamics}
The derivation of the linearized dynamics for the mean spin components ($\langle \hat{f}^{\beta}_{\alpha,x} \rangle$ or $\langle \hat{\breve{f}}^{\beta}_{\alpha,x} \rangle$) is best approached by adopting the method introduced in \cite{PhysRevA.16.1877} (see also the Supplementary material in \cite{katz2015coherent}), where spin operators are expressed as spherical tensor operators in the coupled ($F,m_F$) basis. For transverse spin, it is sufficient to analyze the dynamics of the $\pm1$ components of the rank-1 spherical tensor (with the zero-component defined by the direction of the applied DC magnetic field). We focus on studying the noise correlation occurring at Zeeman frequencies. Hyperfine coherences are not measured, and their contribution to Zeeman dynamics can be neglected in the zero polarization limit.

Here, for the reader's convenience, we provide a concise overview of the contribution of spin-exchange interactions between different alkali species in the dynamics of the mean spins. We refer to \cite{PhysRevA.16.1877} and \cite{katz2015coherent} for more details.

The equation of motion for the density matrix of species $\beta$ due to spin-exchange collision with species $\gamma$ can be written in the form:
\begin{align}
\frac{1}{R_{\text{se}}^{\beta,\gamma}}\frac{d \rho_\beta}{dt}  &= \sum_m \left \{ \sqrt{\frac{[I_{\gamma}]}{[I_\beta]}} \langle T_{001m}^{\gamma \dag}\rangle-\langle T_{0 0 1 m}^{\beta \dag}\rangle \right \}  T_{001m}^{\beta} \nonumber \\
&+\sum_{\substack{\Lambda \mu m \\ \Lambda \neq 0} }  \sqrt{2 [I_\gamma]} \langle T_{\Lambda \mu 00}^{\beta \dag} \rangle \langle T_{001m}^{\gamma\dag} \rangle  T_{\Lambda \mu 1 m}^{\beta} \nonumber \\
&- \rho_\beta +\sum_{\Lambda \mu} T_{\Lambda \mu 00}^{\beta} \langle T_{\Lambda \mu 00}^{\beta \dag} \rangle+ \sum_m \langle T_{0 0 1 m}^{\beta\dag} \rangle T_{0 0 1 m}, \label{eq:densityMatrixEquationSphericalTensors}
\end{align}
where $[I]=2I+1$, and $T_{\Lambda \mu l m} \equiv T_{\Lambda \mu} (I I) \otimes T_{lm}(SS)$ is the spherical tensor spin operator in the uncoupled basis, with $T_{LM}(K K')$ being the spherical tensor in the angular momentum basis with quantum numbers $K$ and $K'$.

The second line in Eq.~\ref{eq:densityMatrixEquationSphericalTensors} has only terms that are proportional to the product of the electron polarization of one species times the nuclear polarization ($\Lambda \neq 0$) of the other species. These terms introduce non-linearity in the atomic evolution since each polarization (electron or nuclear) depends on the density matrix. However, in the case of thermal-unpolarized atomic ensembles, these terms are second order with respect to the  small quantity of polarization and can be neglected; consequently, linearized dynamics can be considered for such systems.

Multiplying both sides of Eq.~\ref{eq:densityMatrixEquationSphericalTensors} with $T_{1M}^{\beta\dag}(FF)$ and taking the trace, the equation of motion for $\langle T_{1M}^{\beta\dag}(FF)\rangle$ can be obtained. The linearized equation contains solely terms of the form $\langle  T_{1M}^{\beta\dag}(FF) T_{0 0 1 m} \rangle$ or $\langle  T_{1M}^{\beta\dag}(FF) T_{\Lambda \mu 00} \rangle$, which can be found by expressing $T_{0 0 1 m}$ and $T_{\Lambda \mu 00}$ in the coupled basis (see Eqs.~63 and 64 in \cite{PhysRevA.16.1877}), and using the orthogonality property for spherical operators,
$\langle T^{\dag}_{LM}(FF') T_{lm}(ff')\rangle = \delta_{Ll}\delta_{Mm}\delta_{Ff}\delta_{F'f'}$. By also employing the definition for the Hermitian conjugate, $T^{\dag}_{LM}(FF') = (-1)^{F-F'+M} T_{L-M}(FF')$, all equations can be readily transformed into equations of motion for the operators $T_{LM}(FF)$.

The contribution of other relaxation mechanisms to the dynamics of mean spins can be determined using similar calculations. Finally, the dynamics due to magnetic field are most easily determined from the Heisenberg equation of motion, while the hyperfine interaction does not affect the Zeeman spherical operators $T_{LM}(FF)$. Equations of motion for the spin density operators $\breve{T}_{LM}(FF)$ can be obtained straightforwardly by utilizing Eqs. \eqref{eq:densityOperatorSingleAtomSpin} and \eqref{eq:densityMatrixEquationSphericalTensors}. This amounts to modifying the rate in the first term of the sum appearing in the first line of the RHS in Eq.~\ref{eq:densityMatrixEquationSphericalTensors}, replacing $R_{\text{se}}^{\beta,\beta'} \propto n_{\beta'}$ with $R_{\text{se}}^{\beta',\beta} \propto n_{\beta}$, i.e. interchanging the spin-exchange rates in the terms of the equations of motion that are responsible for the mixing between the spin operators of the two species. 

Overall, the linearized equations of motion for the spin density operators of the two species can be compactly written as:
\begin{equation}
\frac{d\mathbf{T}}{dt} = 
\begin{pmatrix} \tilde{A} &  \mathbf{0}_4 \\
\mathbf{0}_4 & \tilde{A}^* 
\end{pmatrix}
\mathbf{T}, \label{eq:linearizedAequations}
\end{equation}
where $\mathbf{0}_4$ is the $4 \times 4$ zero matrix and $\mathbf{T}=[\mathbf{T}_{11}  \phantom{\small{a}} \mathbf{T}_{1-1}]^{\top}$, with $\mathbf{T}_{11}= [T_{11}^{\Rb}(aa) \phantom{\small{a}} T_{11}^{\Rb}(bb) \phantom{\small{a}} T_{11}^{\Cs}(a'a')\phantom{\small{a}} T_{11}^{\Cs}(b'b')]$ and similarly for $\mathbf{T}_{1-1}$. We remind the reader that $a=I_\Rb+1/2$ and $b=I_\Rb-1/2$, while $a'=I_\Cs+1/2$ and $b'=I_\Cs-1/2$. The $4\times 4$ matrix $\tilde{A}$ is:
\begin{align}
\tilde{A}  &=  \begin{pmatrix}
R_{\text{se}}^{\Rb,\Rb} \Phi_{\rm{se}}(I_\Rb) & \mathbf{0}_2 \\
 \mathbf{0}_2 & R_{\text{se}}^{\Cs,\Cs}  \Phi_{\rm{se}}(I_{\Cs})
\end{pmatrix} \nonumber\\
&+ \begin{pmatrix}
R_{\text{se}}^{\Rb,\Cs} \tilde{\Phi}(I_\Rb) & R_{\text{se}}^{\Cs,\Rb} \breve{\Phi}^{\top}_{\rm{se}} /[I_\Rb] \\
 R_{\text{se}}^{\Rb,\Cs} \breve{\Phi}_{\rm{se}} /[I_{\Cs}] & R_{\text{se}}^{\Cs,\Rb} \tilde{\Phi}(I_{\Cs})
\end{pmatrix} \nonumber \\
&+ R\begin{pmatrix}
\tilde{\Phi} (I_\Rb) & \mathbf{0}_2 \\
 \mathbf{0}_2 & \tilde{\Phi}(I_{\Cs})
\end{pmatrix} \nonumber\\
&-R_D \begin{pmatrix}
 \mathbf{I}_2 & \mathbf{0}_2 \\
 \mathbf{0}_2 &  \mathbf{I}_2
\end{pmatrix} \nonumber \\
&+i\omega_0 
\begin{pmatrix}
\Phi_{B}(I_\Rb) & \mathbf{0}_2 \\
\mathbf{0}_2 & \Phi_{B}(I_{\Cs})
\end{pmatrix}\label{eq:MatrixEvolutionAtildeDef}
\end{align}
where $\mathbf{0}_2$ and $\mathbf{I}_2$ are respectively the $2\times2$ zero and identity matrices, and
\begin{widetext}
\begin{align}
 \Phi_{\rm{se}}(I)&={1\over {3[I]^2}}
\begin{pmatrix}
-2 I (2 I-1) & 2 \sqrt{I (I+1) (2I-1) (2 I+3)}  \\
 2 \sqrt{I (I+1) (2 I-1) (2I+3)} & -2(I+1) (2 I+3) 
\end{pmatrix}\nonumber\\
\tilde{\Phi}(I)&={1\over {[I]^2}}
\begin{pmatrix}
 -(2 I^2+I+1) & \sqrt{I (I+1) (2 I-1) (2 I+3)} \\
 \sqrt{I (I+1) (2 I-1) (2 I+3)} & -(2 I^2+3 I+2)
\end{pmatrix}\nonumber\\
{\Phi}_{\rm{se}} &= \frac{1}{3\sqrt{[I_\Rb][I_\Cs]}} 
\begin{pmatrix}
 \sqrt{(I_\Rb+1) (2 I_\Rb+3)( I_\Cs+1) (2 I_\Cs+3)} & -\sqrt{I_\Rb (2 I_\Rb-1) (I_\Cs+1) (2 I_\Cs+3)} \\
 -\sqrt{(I_\Rb+1) (2 I_\Rb+3) I_{\Cs} (2 I_\Cs-1)} &  \sqrt{I_\Rb (2 I_\Rb-1) I_{\Cs} (2 I_\Cs-1)} 
\end{pmatrix}\nonumber\\
\Phi_{B}(I) &={1\over {[I]}}
\begin{pmatrix}
1& 0 \\
0 & -1
\end{pmatrix}
\end{align}
\end{widetext}
In Eq.~\ref{eq:MatrixEvolutionAtildeDef}, we took into account that the S-damping relaxation rate, $R$, and the relaxation rate due to diffusion, $R_D$, is approximately the same for both species.

Converting Eq.~\ref{eq:linearizedAequations} to an equation of motion for the mean transverse Cartesian components of the spin-density operators simply involves a change of basis by a similarity transformation. Defining the vector $\mathbf{X} = [\breve{f}_{a,x}^\Rb,\breve{f}_{b,x}^\Rb,\breve{f}_{a,y}^\Rb,\breve{f}_{b,y}^\Rb,\breve{f}_{a',x}^\Cs,\breve{f}_{b',x}^\Cs,\breve{f}_{a',y}^\Cs,\breve{f}_{b',y}^\Cs ]^\top$, the dynamical evolution of the mean is given by:
\begin{equation}
\frac{d \langle \mathbf{X} \rangle}{dt} = A \langle \mathbf{X} \rangle, \label{eq:ap:MeanXTimeEvolutionA}
\end{equation}
where $A=\mathcal{M} \tilde{A} \mathcal{M}^{-1}$. The matrix $\mathcal{M}$ represents the change of basis: $\mathbf{X}=\mathcal{M} \mathbf{T}$ and has the form:
\begin{equation}
\mathcal{M} = \begin{pmatrix}
-\mathfrak{t} & \mathfrak{t} \\
i \mathfrak{t} & i \mathfrak{t}
\end{pmatrix},
\end{equation}
\begin{equation}
\mathfrak{t} = \begin{pmatrix}
\mathfrak{t}_{1}(I_\Rb) & 0 & 0 & 0 \\
 0 & \mathfrak{t}_{2}(I_\Rb) & 0 & 0 \\
 0 & 0 &  \mathfrak{t}_{1}(I_{\Cs})  & 0 \\
 0 & 0 & 0 & \mathfrak{t}_{2}(I_{\Cs})  
\end{pmatrix},
\end{equation}
where $\mathfrak{t}_{1}(I)={\sqrt{(I+1) (2 I+1) (2I+3)}}/{2 \sqrt{3}}$ and $\mathfrak{t}_{2}(I)={\sqrt{I (2 I-1) (2 I+1)}}/{2 \sqrt{3}}$.

We note that $A$ is a diagonalizable matrix with only real elements and can be written in the block form:
\begin{equation}
A=
\begin{pmatrix}
\mathcal{B} & \mathcal{D} \\
-\mathcal{D} & \mathcal{B}
\end{pmatrix}, \label{eq:AMatrixBlockDiagonalForm}
\end{equation}
where $\mathcal{B}$ and $\mathcal{D}$ are $4 \times 4$ square matrices, with the diagonal matrix $\mathcal{D}$ describing the coupling to the DC magnetic field. The physical meaning underlying this structure of $A$ is that the two transverse spins share identical dynamics.


\section{\label{sec:AppC}Properties of the covariance matrix}

\subsection{General properties} \label{sec:ap:RGeneralProp}
Here, we derive some useful properties of the covariance matrix. We define the vector $\mathbf{\hat{X}}(t)=\left[\hat{x}_1(t), \hat{x}_2(t), ... \right]^{T}$, where $\hat{x}_i(t)$ are  Hermitian operators (in our experiment these are Cartesian hyperfine spin operators) whose mean evolves in time according to Eq.~\ref{eq:ap:MeanXTimeEvolutionA}. The symmetrized covariance matrix has components $R_{ij}(\tau) = \langle \hat{x}_i(t+\tau) \hat{x}_j(t)+\hat{x}_j(t) \hat{x}_i(t+\tau) \rangle/2$. The dependence on $\tau$ can be found from the regression theorem (see section IV.C) to be \cite{gardiner2009stochastic}:
\begin{equation}
R(\tau) = 
\begin{dcases}
e^{A \tau} R(0) &, \tau \geq 0 \\
R(0) e^{-A^{\top} \tau} &, \tau <0,
\end{dcases}. \label{eq:ap:CovarianceEvolution0}
\end{equation}

First, we notice that the stationary ($\tau=0$) covariance matrix is symmetric: $R^\top(0) = R(0)$. For the transpose of the covariance matrix we find ($\tau \geq 0$):
\begin{equation}
R^\top(\tau) = R^\top(0) e^{A^\top\tau} \Rightarrow R^\top(\tau) = R(0) e^{A^\top|\tau|} = R(-\tau),
\end{equation}
i.e., $R_{ij}(\tau) = R_{ji}(-\tau)$ as should be expected for any stationary process. 

The spectrum matrix is (assuming the measurement time to be infinite in Eq.\eqref{eq:psd_SA}):
\begin{equation}
S(\omega) = \frac{1}{2 \pi} \int_{-\infty}^{+\infty} R(\tau) e^{-i \omega \tau} d \tau.
\label{eq:psd}
\end{equation}
The reality of $R(\tau)$ yields $S^{*}(\omega) = S(-\omega)$. Taking into account that $R_{ij}(\tau) = R_{ji}(-\tau)$ we obtain: 
\begin{align}
S^{*}_{ij}(\omega) & =  \frac{1}{2 \pi} \int_{-\infty}^{+\infty} R_{ij}(\tau) e^{i \omega \tau} d \tau \\
& \stackrel{\tau\rightarrow -\tau}{=}\frac{1}{2 \pi} \int_{-\infty}^{+\infty} R_{ij}(-\tau) e^{-i \omega \tau} d \tau \\
& = \frac{1}{2 \pi} \int_{-\infty}^{+\infty} R_{ji}(\tau) e^{-i \omega \tau} d \tau = S_{ji}(\omega).
\end{align}
The spectrum can also be written in the form:
\begin{align}
S_{ij}(\omega) & = \int_{-\infty}^{+\infty} d\tau R_{ij}(\tau) e^{-i \omega \tau} \\
& = \int_{-\infty}^{0} d\tau R_{ij}(\tau) e^{-i \omega \tau}+ \int_{0}^{+\infty} d\tau R_{ij}(\tau) e^{-i \omega \tau} \\
&=  \int_{0}^{+\infty} d\tau R_{ij}(-\tau) e^{i \omega \tau}+\int_{0}^{+\infty} d\tau R_{ij}(\tau) e^{-i \omega \tau} \\
&= \int_{0}^{+\infty} d\tau R_{ji}(\tau) e^{i \omega \tau}+\int_{0}^{+\infty} d\tau R_{ij}(\tau) e^{-i \omega \tau}.
\end{align}

In general, $S_{ij}(\omega)$ is complex, with real and imaginary parts given by 
\begin{align}
\operatorname{Re} \left[ S_{ij}(\omega) \right] &=\int_{0}^{+\infty} d\tau  \left[ R_{ij}(\tau)+R_{ji}(\tau)\right] \cos(\omega \tau)\nonumber\\
\operatorname{Im} \left[ S_{ij}(\omega) \right] &= \int_{0}^{+\infty} d\tau  \left[ R_{ij}(\tau)-R_{ji}(\tau)\right] \sin(\omega \tau).\nonumber
\end{align}
If the spectrum is real we find:
\begin{equation}
\begin{split}
S_{ij}(\omega) = S_{ij}^{*}(\omega) \Rightarrow R_{ij}(\tau) = R_{ji} (\tau) =R_{ij} (-\tau). \label{eq:Ap:RealSpectrumSymR}
\end{split}
\end{equation}
The reverse is also true, $R_{ij}(\tau) = R_{ji} (\tau)$ implies $S_{ij}(\omega)$ is real.

An interesting case, relevant for spin-noise measurements, arises when the stationary covariance matrix is diagonal ($R_{ij}(0) = \Delta x^2_{i} \delta_{ij}$, $\Delta x^2_{i}$ being the variance of $\hat{x}_i$), and the spectrum is real, i.e. the covariance matrix is symmetric (see Eq.~\eqref{eq:Ap:RealSpectrumSymR}). In this case, a condition is imposed between the noise variance terms and the matrix $A$:
\begin{align}
& R_{ij}(\tau) = R_{ji}(\tau)  \Rightarrow \sum_k \left[ e^{A \tau}\right]_{ik} R_{kj}(0)  = \sum_k \left[ e^{A \tau}\right]_{jk} R_{ki}(0) \\
& \xRightarrow[ ]{R_{ij}(0)  = \Delta x^2_{i} \delta_{ij} } \left[ e^{A \tau}\right]_{ij} \Delta x^2_{j} = \left[ e^{A \tau}\right]_{ji} \Delta x^2_{i} \nonumber\\
&\Rightarrow \frac{ \left[ e^{A \tau}\right]_{ij}}{ \left[ e^{A \tau}\right]_{ji}} = \frac{\Delta x^2_{j}}{\Delta x^2_{i}}. \label{eq:App:DiagCovRealSpectrum}
\end{align}
Here, $[..]_{ij}$ denotes the ${ij}$ element of the matrix in the square brackets.

We note that if the steady-state spin-covariance matrix is diagonal, symmetry arguments dictate that the spin-noise in each hyperfine level scales as:
\begin{equation}
 \Delta x^2 \propto n f(f+1) \times (2f+1)/(2I+1), \label{eq:Ap:NoiseScaling}
\end{equation}
where $n$ is the atomic density, $f$ is quantum number of total atomic spin in the given hyperfine state and $I$ is the nuclear spin. As shown below, for this type of noise, the spin dynamics satisfy the constraint expressed in \ref{eq:App:DiagCovRealSpectrum}.

\subsection{Properties for spin dynamics}
In particular for the spin dynamics presented in \ref{sec:MeanSpinDynamics}) and captured in matrix $A$, the rules of matrix multiplication can be used to show that all the powers of $A$ and consequently the matrix exponential $e^{A \tau}$ has a similar to Eq.~\ref{eq:AMatrixBlockDiagonalForm} pattern, that is:
\begin{equation}
e^{A \tau} = \sum_{n=0}^{\infty} \frac{A^n \tau^n}{n!} = \begin{pmatrix}
\mathfrak{B} & \mathfrak{D} \\
-\mathfrak{D} & \mathfrak{B} \\
\end{pmatrix}, \label{eq:Ap:BlockMatrixForm}
\end{equation}
though in this case $\mathfrak{D}$ is not diagonal. This structure expresses the fact that the two transverse spin components ($x$ and $y$ here) are physically equivalent.

The matrix elements of $\mathfrak{B}$ satisfy the Equation:
\begin{equation}
\frac{\left[ \mathfrak{B}\right]_{ij}}{\left[ \mathfrak{B}\right]_{ji}} = [r]_{ij}, \label{eq:ap:ratioBelements}
\end{equation}
where:
\begin{widetext}
\begin{equation}
r=
\begin{pmatrix}
 1 & \frac{(I_1+1) (2 I_1+3)}{I_1 (2
   I_1-1)} & \frac{(I_1+1) (2 I_1+3)
   R_{\rm{se}}^{21}}{(I_2+1) (2 I_2+3)
   R_{\rm{se}}^{12}} & \frac{(I_1+1) (2 I_1+3)
   R_{\rm{se}}^{21}}{I_2 (2 I_2-1) R_{\rm{se}}^{12}}
   \\
 \frac{I_1 (2 I_1-1)}{(I_1+1) (2
   I_1+3)} & 1 & \frac{I_1 (2 I_1-1)
   R_{\rm{se}}^{21}}{(I_2+1) (2 I_2+3)
   R_{\rm{se}}^{12}} & \frac{I_1 (2 I_1-1)
   R_{\rm{se}}^{21}}{I_2 (2 I_2-1) R_{\rm{se}}^{12}}
   \\
 \frac{(I_2+1) (2 I_2+3)
   R_{\rm{se}}^{12}}{(I_1+1) (2 I_1+3)
   R_{\rm{se}}^{21}} & \frac{(I_2+1) (2 I_2+3)
   R_{\rm{se}}^{12}}{I_1 (2 I_1-1) R_{\rm{se}}^{21}}
   & 1 & \frac{(I_2+1) (2 I_2+3)}{I_2
   (2 I_2-1)} \\
 \frac{I_2 (2 I_2-1)
   R_{\rm{se}}^{12}}{(I_1+1) (2 I_1+3)
   R_{\rm{se}}^{21}} & \frac{I_2 (2 I_2-1)
   R_{\rm{se}}^{12}}{I_\beta (2 I_1-1) R_{\rm{se}}^{21}}
   & \frac{I_2 (2 I_2-1)}{(I_2+1) (2
   I_2+3)} & 1 
\end{pmatrix},
\end{equation}
\end{widetext}
where for a lighter notation we made the substitution $\Rb\rightarrow 1$ and $\Cs\rightarrow 2$.
To prove this, it is shown through mathematical induction that the block diagonal matrices corresponding to the various integer powers of $A$ adhere to the condition outlined in Equation~\ref{eq:ap:ratioBelements}. 

The covariance matrix takes the form:
\begin{equation}
R(\tau) =
\begin{pmatrix}
\mathfrak{B} \Sigma_{xx}+ \mathfrak{D} \Sigma_{yx} & \mathfrak{B} \Sigma_{xy}+ \mathfrak{D} \Sigma_{yy} \\
\mathfrak{B} \Sigma_{yx}- \mathfrak{D} \Sigma_{xx} & \mathfrak{B} \Sigma_{yy}-\mathfrak{D} \Sigma_{xy}
\end{pmatrix}, \label{eq:ap:RtauConciseForm}
\end{equation}
where the $4\times 4$ matrices $\Sigma_{xx}$, $\Sigma_{yy}$, $\Sigma_{xy}$, $\Sigma_{yx}$ (with the $x$, $y$ subscripts indicating Cartesian components) are defined from the following equation:
\begin{equation}
R(0)=\begin{pmatrix}
\Sigma_{xx} & \Sigma_{xy} \\
\Sigma_{yx} & \Sigma_{yy}
\end{pmatrix}. \label{eq:ap:SigmaMatricesDef}
\end{equation}
The symmetry of $R(0)$ dictates that $\Sigma_{xy}=\Sigma_{yx}^{\top}$. Furthermore, the physical condition of identical noise behavior \textit{at all times} for the transverse Cartesian components gives $\Sigma_{xx} = \Sigma_{yy}$ and (see Eq.~\ref{eq:ap:RtauConciseForm}):
\begin{equation}
\mathfrak{B} \Sigma_{xx}+ \mathfrak{D} \Sigma_{yx} =  \mathfrak{B} \Sigma_{yy}-\mathfrak{D} \Sigma_{xy} \Rightarrow  \Sigma_{yx}+\Sigma_{xy}=0. \label{eq:ap:SymmetryOfSigmaxy}
\end{equation}

For unpolarized thermal atoms it is natural to assume that the equal time covariance between the different transverse spins is symmetric with respect to the interchange of the transverse components, i.e. $\Sigma_{xy}$ (and similarly  $\Sigma_{yx} = \Sigma_{xy}^{T}$) is symmetric:  $\Sigma_{xy}= \Sigma_{yx}^{\top}$. Augmenting this with Eq.~\eqref{eq:ap:SymmetryOfSigmaxy} we find that $\Sigma_{xy} = \Sigma_{yx} =0$, and the covariance matrix ($\tau \geq 0$) takes the simpler form
\begin{equation}
R(\tau) =
\begin{pmatrix}
\mathfrak{B} \Sigma & \mathfrak{D} \Sigma \\
- \mathfrak{D} \Sigma & \mathfrak{B} \Sigma
\end{pmatrix}, \label{eq:ap:RtauConciseFormN}
\end{equation}
where $\Sigma=\Sigma_{xx}=\Sigma_{yy}$.

We now consider the case where there are no equal-time correlations between the different spin components, i.e. $\Sigma$ is diagonal. As discussed above (see Eq.~\ref{eq:Ap:NoiseScaling}), in this case the noise variance of the collective spin in the thermal state scales according to the expected number of atoms in the corresponding hyperfine state and the magnitude of the spin in this state. Combining this with Eq.~\ref{eq:ap:ratioBelements}, we find that the covariance block matrix $R_{xx}(\tau)=R_{yy}(\tau)=\mathfrak{B} \Sigma$ is symmetric:
\begin{equation}
\begin{split}
\left[ \mathfrak{B} \Sigma \right]_{ij} & = \sum_{k}\left[ \mathfrak{B} \right]_{ik} \delta_{kj} \left[ \Sigma \right]_{jj} = \left[ \mathfrak{B} \right]_{ij}  \left[ \Sigma \right]_{jj} \\
& \stackrel{Eq. \eqref{eq:ap:ratioBelements}}{=} \left[ \mathfrak{B} \right]_{ji}  \left[ \Sigma \right]_{ii} = \left[ \mathfrak{B} \Sigma \right]_{ji}.
\end{split}
\end{equation}
This symmetry implies that the cross-spectrum between spin components in the same transverse (Cartesian) axis is strictly real, i.e. the imaginary part is zero (see discussion in \ref{sec:ap:RGeneralProp}).

Inversely,  if all the cross-spectra between the 4 spins (2 hyperfine spins for each of the 2 species) are strictly real, then it can be proven that the equal time covariance matrix $\Sigma$ is symmetric, under the physically justifiable assumption that the variances for the various spin components scale according to Eq.~\eqref{eq:Ap:NoiseScaling}. In the following we sketch the proof of the above statement. The reality of the spectra implies that the block matrix $\mathfrak{B} \Sigma$ is symmetric: 
\begin{equation}
\mathfrak{B} \Sigma = \Sigma^{\top} \mathfrak{B}^{\top} \label{eq:ap:BSsymmetryCondition}
\end{equation}
Considering that $\Sigma=\Sigma^{\top}$ and the assumption for the scaling of spin variances (Eq.~\eqref{eq:Ap:NoiseScaling}), there remain 6 elements (i.e. the elements located above -or lower- the main diagonal) to be determined from the system of equations introduced in Eq.~\ref{eq:ap:BSsymmetryCondition}. 
Under general conditions, given that $\mathfrak{B}$ satisfies the condition in Eq.~\ref{eq:ap:ratioBelements}, the system of equations becomes non-singular for these 6 elements. Consequently, solving the system reveals that these elements are all zero.

\section{\label{sec:LightShift}Light-shift noise}
Here, on qualitative grounds we argue that the light-shift noise induced onto the equilibrium atomic ensemble from the probe beam (also termed back-action noise \cite{optical_mag}) is significantly smaller than the spin-noise and can therefore be ignored. The Hamiltonian describing the probe light-shift experienced by an atom in a particular hyperfine state is: $H_{\text{LS}} \propto \hat{S}_3 (t)\hat{F}_x (t)$, where $\hat{F}_x(t)$ is the hyperfine angular momentum in the direction of probe propagation and $\hat{S}_3(t)$ is the Stokes element measuring the flux difference between the right- and left- circular components, which for linearly polarized probe only describes the quantum polarization fluctuations.  This Hamiltonian is formally equivalent to the coupling of a (white-noise) transverse magnetic field to the atoms. The rms amplitude of the effective magnetic field is at the \si{\femto\tesla} level for typical experimental conditions \cite{PhysRev.171.11}. Clearly, the insensitivity of unpolarized atomic ensembles to magnetic-fields also applies to the light-shift noise. 

\section{\label{sec:app:BandwidthEffect}Bandwidth effect on the observed noise}
We here provide a brief explanation of the bandwidth effect on the recorded noise spectrum. First, we consider the power when digitizing a signal with a finite sampling rate. While the precise relationship between the recorded data point ($\tilde{y}$) and the underlying actual analog signal $y$ may slightly vary depending on the data acquisition system, we model this relationship as: $\tilde{y}(t) = \frac{1}{\Delta T}\int_t^{t+\Delta T} y(t') dt'$, where $1/\Delta T$ represents the sampling rate. In this case, the zero-time cross-correlation power can be expressed as
\begin{align}
\langle \tilde{y}_a \tilde{y}_b \rangle & = \frac{1}{\Delta T^2}\int_t^{t+\Delta T} dt' \int_t^{t+\Delta T} dt'' \langle y_a(t') y_b(t'') \rangle \label{eq:App:BWEffect1}\\
& = \frac{1}{\Delta T^2}\int_{t}^{t+\Delta T} dt' \int_{t'-t-\Delta T}^{t'-t} d\tau R_{ab}(\tau) \label{eq:App:BWEffect2} \\
& = \frac{1}{\Delta T^2}\int_{t}^{t+\Delta T} dt' \int_{t'-t-\Delta T}^{t'-t} d\tau \int_{-\infty}^{\infty } d\omega S_{ab}(\omega) e^{\imath \omega \tau}  \label{eq:App:BWEffect3} \\
& = \int_{-\infty}^{\infty} S_{ab}(\omega) \left[  \frac{\sin \left( \frac{\omega \Delta T}{2} \right)}{\frac{\omega \Delta T}{2}}\right]^2 d \omega, \label{eq:App:BWEffect4}
\end{align}
where $S_{ab} (\omega) = \frac{1}{2 \pi } \int_{-\infty}^{\infty} R_{ab}(\tau) e^{-\imath \omega \tau}$ is the cross-correlation spectrum, and $R_{ab}$ is the correlation between $y_a$ and $y_b$. In passing from Eq.\eqref{eq:App:BWEffect1} to Eq.\eqref{eq:App:BWEffect2} we took into account that for stationary processes the correlation $\langle y_a(t')y_b(t'')\rangle$ only depends on the time difference $t'-t''$ and performed a change of variables: $(t',t'') \rightarrow (t',\tau =t'-t'')$. If the two signals are filtered with the same filter (e.g. an anti-aliasing filter) then Eq.\eqref{eq:App:BWEffect4} is modified to:
\begin{equation}
    \langle \tilde{y}_a \tilde{y}_b \rangle =  \int_{-\infty}^{\infty} S_{ab}(\omega) \left[  \frac{\sin \left( \frac{\omega \Delta T}{2} \right)}{\frac{\omega \Delta T}{2}}\right]^2 \Phi (\omega) d \omega, \label{eq:App:BWEffect41}
\end{equation}
where $\Phi (\omega)$ is the filter (in power).

We now examine the noise-power of lock-in amplifier signals. For concreteness, we consider a simplified version of a lock-in amplifier signal $\tilde{y}$, given from: $\tilde{y} = (1/T_{\text{BW}}) \int_{0}^{T} y(t') ~\mathrm{exp}[-(T-t')/T_{\text{BW}}]\cos (\omega_0 t') d t'$, where $\omega_0$  is the demodulation frequency, $T_{\text{BW}}$ is the lock-in time-constant being effectively the inverse of the measurement bandwidth (assuming 3 dB roll-off), $y(t)$ is the actual signal, and $T$ is the measurement time. The zero-time correlation power between two such lock-in signals is:
\begin{widetext}
\begin{align}
\langle \tilde{y}_a \tilde{y}_b \rangle & =  \frac{1}{T_{\text{BW}}^2}\int_0^{T} dt' \int_0^{T} dt'' \langle y_a(t') y_b(t'') \rangle e^{-\frac{T-t'}{T_{\text{BW}}}} e^{-\frac{T-t''}{T_{\text{BW}}}} \cos (\omega_0 t') \cos (\omega t'') \label{eq:App:BWEffect5}\\
&=   \frac{1}{T_{\text{BW}}^2}\int_0^{T} dt' \int_{t'-T}^{t'} d\tau R_{ab}(\tau) e^{-\frac{T-t'}{T_{\text{BW}}}} e^{-\frac{T-(t'-\tau)}{T_{\text{BW}}}} \cos(\omega_0 t') \cos \left[ \omega_0 (t'-\tau) \right] \label{eq:App:BWEffect6} \\
& =   \frac{1}{T_{\text{BW}}^2}\int_{-\infty}^{\infty} d \omega S_{ab} (\omega) \int_0^{ T} dt' e^{-2\frac{T-t'}{T_{\text{BW}}}} \cos(\omega_0 t') \int_{t'-T}^{t'}   d\tau   e^{\frac{\tau}{T_{\text{BW}}}} \cos \left[ \omega_0 (t'-\tau) \right] e^{\imath \omega \tau} = \int_{-\infty}^{\infty} d \omega S_{ab} (\omega) \phi(\omega), \label{eq:App:BWEffect7} \\
&\phi(\omega) = \frac{\frac{1}{T_{\text{BW}}}^2 \left[ \frac{1}{T_{\text{BW}}}^2+\left(\frac{1}{T_{\text{BW}}}^2+\omega^2-\omega_0^2\right) \cos (2 T \omega_0)+2\frac{1}{T_{\text{BW}}}  \omega_0 \sin (2 T \omega_0)+\omega^2+\omega_0^2\right]}{2 \left [ \left(\frac{1}{T_{\text{BW}}}^2+\omega^2\right)^2+2 \omega_0^2 (\frac{1}{T_{\text{BW}}} -\omega ) (\frac{1}{T_{\text{BW}}} +\omega )+\omega_0^4\right]} \label{eq:App:BWEffect8} \\
& \approx \frac{1}{4} \frac{(1/T_{\text{BW}})^2}{(1/T_{\text{BW}})^2+(\omega-\omega_0)^2}. \label{eq:App:BWEffect9}
\end{align}
\end{widetext}
The filter function $\phi (\omega)$ was calculated in the (stationary) limit $T\gg T_{\text{BW}}$. The approximation in Eq.~\eqref{eq:App:BWEffect9} holds for $\omega_0 T_{\text{BW}} \gg 1 $ and $|\omega_0-\omega | \ll \omega_0$. 

\bibliography{references}

\begin{thebibliography}{127}
\bibitem [{com()}]{comParaxModes}%
  \BibitemOpen
  \href@noop {} {}\bibinfo {note} {Here, the paraxial scattering in spatial
  modes other than the probe mode is neglected. For a typical experiment with
  warm atoms contained inside a glass cell where the spatial profile of the
  probe beam is smaller than the size of the atomic ensemble, this type of
  scattering is negligible. See \cite{PhysRevA.89.033850} for
  details.}\BibitemShut {Stop}%
\bibitem [{not()}]{note:ThickEnsemble}%
  \BibitemOpen
  \href@noop {} {}\bibinfo {note} {To show this, the ensemble is divided into
  thin layers along the direction of probe propagation. In each layer, the
  light undergoes optical rotation: $d\phi = \bar{g} d \hat{F}/2$. Absorption
  is assumed to be equal for the $y$ and $z$ component (i.e. linear dichroism
  is negligible) and is modelled as a typical lossy channel introducing vacuum
  noise.}\BibitemShut {Stop}%
\end{thebibliography}

\end{document}